\documentclass[aps,prb,superscriptaddress,twocolumn]{revtex4-1}

\usepackage[caption=false]{subfig}  
\usepackage{amsmath}
\usepackage{tikz}
\usepackage{float}
\usepackage{flafter}
\usetikzlibrary{shapes}
\usepackage{graphicx}

\begin{document}

\title{Anisotropy of electronic stopping power in graphite}

\author{Jessica Halliday}
\affiliation{Theory of Condensed Matter,
             Cavendish Laboratory, University of Cambridge, 
             J. J. Thomson Ave, Cambridge CB3 0HE, United Kingdom}
             
\author{Emilio Artacho}
\affiliation{Theory of Condensed Matter,
             Cavendish Laboratory, University of Cambridge, 
             J. J. Thomson Ave, Cambridge CB3 0HE, United Kingdom}
\affiliation{CIC Nanogune and DIPC, Tolosa Hiribidea 76, 
             20018 San Sebastian, Spain}
\affiliation{Ikerbasque, Basque Foundation for Science, 48011 Bilbao, Spain}

\date{\today}

\begin{abstract}
  The rate of energy transfer from ion projectiles onto the electrons of a solid 
target is hard to determine experimentally in the velocity regime between 
the adiabatic limit and the Bragg peak. 
  First-principles simulations have lately offered relevant new insights and 
quantitative information for prototypical homogeneous materials.
  Here we study the influence of structural anisotropy on electronic stopping 
power with time-dependent density functional theory simulations of a 
hydrogen projectile in graphite.
  The projectile traveled at a range of angles and impact parameters for 
velocities between 0.1 and 1.4 a.u., and the electronic stopping power 
was calculated for each simulation.
  After validation with average experimental data, the anisotropic crystal 
structure was found to have a strong influence on the stopping power, 
with a difference between simulations parallel and perpendicular to the 
graphite plane of up to 25\%, more anisotropic than expected based on 
previous work.
  The velocity dependence at low velocity displays clear linear behavior in 
general, except for projectiles traveling perpendicular to graphitic layers, 
for which a threshold-like behavior is obtained.
  For projectiles traveling along graphitic planes metallic behavior is 
observed with a change of slope when the projectile velocity reaches the
Fermi velocity of the electrons.

\end{abstract}

\pacs{PACS: }

\maketitle

\section{Introduction}

  Stopping power is the rate of energy loss along the path of a charged 
particle as it passes through matter.
  Stopping power is of interest in a wide range of areas, from nuclear 
power generation to medical applications \cite{was,cancer}.
  Two mechanisms of energy loss are involved: nuclear stopping, due 
to the interaction of the projectile with the nuclei of the target, and 
electronic stopping, from the interaction of the projectile with the 
electrons of the target.
  This work focusses on the electronic stopping power ($S_{e}$), which 
dominates at high projectile velocities.

 Most of modern electronics is based on materials and heterostructures grown along well-controlled  crystalline directions, a trend that appears to continue in the brave new world of two-dimensional materials, heterostructures and devices based on them. A good characterisation of the orientation dependence of radiation effects would appear 
quite relevant for ascertaining on resilience of such devices. Of particular interest is the velocity regime in which projectiles are fast enough for non-adiabatic effects to be important, but not so fast so as to become insensitive the structure, pointing to the scale of a few tenths of an atomic unit of velocity (1 a.u.$ = c/137$).
  Previous studies in various materials have investigated anisotropy in $S_{e}$ in thin films, particularly at higher energies \cite{doi:10.1080/00337577208231165, PhysRevA.50.4979, PhysRevB.55.4332} However knowledge of the effect of structural anisotropy at velocities of a few tenths of a.u. is still very limited given the difficulties in obtaining experimental information in this range\cite{kaferbock, softky, seltzer, crawford2, PhysRevB.93.035128}.

  Experimentally $S_{e}$ is difficult to measure directly, particularly 
at low velocities where nuclear stopping is also significant; simulations, 
however, allow $S_{e}$ to be directly accessed.
  Echenique $et\ al.$ \cite{echenique81, echenique86} used density 
functional theory to calculate electronic stopping power in jellium, 
capturing non-linear effects and replicating experimental results not 
captured in linear-response theory calculations \cite{ferrell, lindhard1954},
also giving rise to derived simulations, including the 
use of first principles techniques for the indirect calculation of $S_{e}$ 
(for reviews see \cite{ECHENIQUE1990229, sigmund2014}).
   Direct, real-time simulations of the electronic stopping process 
have also been performed during the last decade using a time-dependent 
tight binding description of the electronic structure and dynamics, coupled 
to nuclear dynamics within an Ehrenfest approach \cite{mason2007, race2010}, 
providing interesting and rich qualitative insights, especially powerful 
given the large system size and long time scale affordable with an empirical 
tight-binding scheme.
   In recent years time-dependent density functional theory (TDDFT) 
 has been used to investigate stopping power from first principles in bulk 
 materials of different kinds (metals, semiconductors, insulators)  \cite{krasheninnikov1, pruneda, ahsanzeb2, correa, ahsanzeb, Puska, 
 caro2015, schleife, ullah, caro2017, PhysRevB.85.235435, refId0, PhysRevB.96.115134}. 
 First principles simulation of electronic stopping has successfully 
reproduced many experimental features not captured by other theoretical 
or simulation methods. 

  A prototypical material with a highly anisotropic layered structure 
is graphite, composed of weakly bonded layers of strongly hexagonally 
bonded carbon atoms, resulting in a high degree of inhomogeneity 
in many properties \cite{chung2002}.
  A major use of graphite is as a moderator in the nuclear power industry, 
to absorb and slow down the neutrons generated by the nuclear fission 
processes, in order to control the rate of fission within a nuclear reactor. 
The stopping power of graphite is thus of intrinsic interest, in 
addition to its position as a simple and strongly anisotropic material, 
and was therefore chosen as the target material in this work.
  The velocity dependence of $S_{e}$ is addressed in the work, with 
an emphasis on its variation with trajectory. 

  A previous work on anisotropy of stopping power in graphite is a 
theoretical study by Crawford \cite{crawford2}, using linear response 
theory based on the Cazaux model \cite{Cazaux1970545} for the optical 
constants of graphite parallel and perpendicular to the graphitic layers.
  Crawford's work calculated, for higher energy projectiles, a similar 
relationship between the incident angle of the projectile and the $S_{e}$ 
as this work.
  They found a small anisotropy of $S_{e}$, with a variability of around 
10\% for projectile velocities between 2 and 20 a.u..
  More recent work by Shukri, Bruneval and Reining \cite{PhysRevB.93.035128} used linear response TDDFT to predict the random electronic stopping power in various materials. They found a similar small anisotropy of up to 3\% between the $S_{e}$ along the in-plane and out-of-plane axes of graphite, for velocities between 0 and 4 a.u.. 
  Previous experimental work by Yagi $et\ al.$ \cite{YAGI20049} 
was unable to direct projectiles between the graphitic layers, 
illustrating the usefulness of simulations to investigate the structural 
anisotropy of $S_{e}$.

\begin{figure}
\begin{center}
\subfloat[]{
\includegraphics{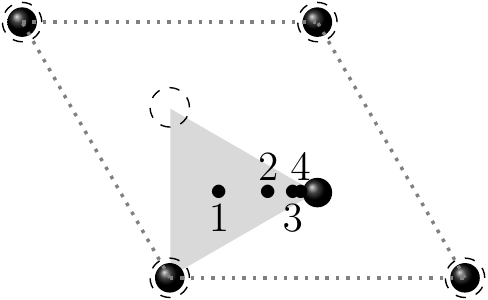}
 \label{1a}}
  \subfloat[]{
\includegraphics{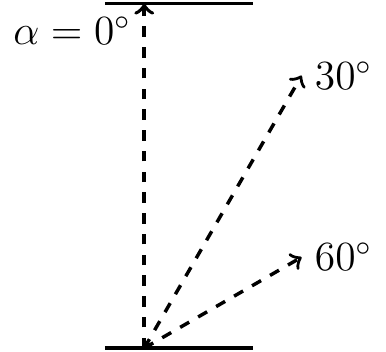}
\label{1b}}
\end{center}
\caption{ (a) Graphite unit cell showing the projectile initial positions 1-4 and 
 (b) trajectories for simulations of projectiles moving out of the plane of graphitic layers. 
 The shaded triangle in (a) is the region of crystallographically unique positions. 
 $\alpha$ is the angle of the trajectory from the graphite $c$ axis, with $\alpha = 0^{\circ}$ 
 perpendicular to the graphitic planes.}
\end{figure}

\section{Method}
\label{method}

\subsection{Simulation details}
 
  Simulations were carried out using the real-time TDDFT implementation 
\cite{tsolakidis, ullah2019} of the SIESTA method \cite{siesta,0953-8984-20-6-064208}.
  The Kohn-Sham orbitals are expanded in a finite basis set of numerical atomic 
orbitals, with the valence electrons of graphite and the projectile represented 
by a double-$\zeta$ polarised basis set.
  The core electrons have been replaced by norm-conserving pseudopotentials 
using the Troullier-Martins scheme \cite{atom, troullier}. Core electrons are known not to interact in stopping processes at low velocities \cite{PhysRevB.93.035128,PhysRevLett.121.116401}.
  The details of both the basis set and the pseudopotentials are specified in the Appendix.
  The local density approximation (LDA) was used for the exchange-correlation 
functional evaluation using the Ceperley-Alder results for the homogeneous 
electron liquid \cite{ceperley}, in the parameterisation of Perdew and 
Zunger \cite{PhysRevB.23.5048}, considering adiabatic time dependence of 
the exchange-correlation functional.

  The ground state of the system is calculated with the projectile stationary 
at its initial position in the graphite box.
  Subsequent TDDFT simulations evolve the electronic wavefunctions according 
to the time-dependent Kohn-Sham equation \cite{kohnsham, runge} as the 
projectile moves at a constant velocity through the box. 
  The forces on all atoms are held as zero throughout the time-dependent 
simulation, so that energy transfer only takes place through inelastic scattering 
to the system electrons.
  This prevents any contribution from nuclear stopping and enables the $S_{e}$ 
to be directly calculated at a single velocity for each simulation.
  The electronic stopping is the average gradient of the total energy of the electronic 
system as a function of the path length of the projectile \cite{ullah}. 
  The error bars in the $S_{e}$ presented in the figures refer to uncertainty 
in fitting to the slope.

\begin{figure}[b]
\begin{center}
\subfloat[]{
\includegraphics{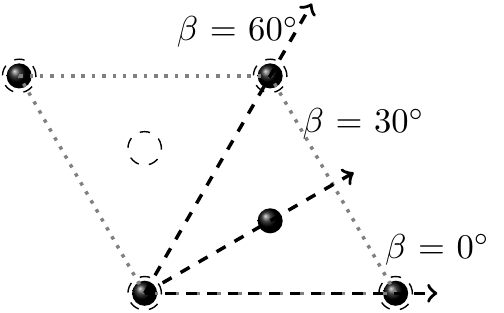}
\label{2a}
}
\subfloat[]{
\includegraphics{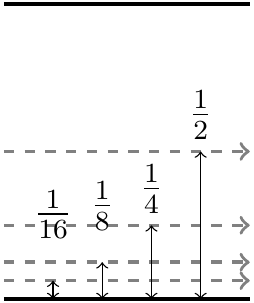}
\label{2b}}
\end{center}
\caption{(a) Graphite unit cell showing the projectile trajectories for 
simulations of projectiles moving parallel to the graphitic layers 
($\alpha=90^{\circ}$). $\beta$ is the angle of the trajectory from the 
$a$ axis of the unit cell. 
  $\beta = 0^{\circ}$ and $60^{\circ}$ are crystallographically equivalent. 
  (b) Distance of projectile trajectories from the graphitic planes.}
\end{figure}

  Projectiles moved the full length of the simulation cell, 13.4 \AA.
  Simulations were carried out using projectiles with velocities between 0.1 and 
1.4 a.u.\cite{PIETSCH197679}. 
  The time-dependent Kohn-Sham equations were integrated using a 
Crank-Nicholson integrator as in Ref.~\onlinecite{tsolakidis} adapted to the 
changing basis and Hilbert space by using a L\"owdin transformation as 
proposed by Sankey and Tomkoff \cite{doi:10.1002/1521-3951} and analysed
in Ref.~\citenum{PhysRevB.95.115155}.
See the Appendix for further details of convergence testing.

\subsection{Simulation trajectories}

  Simulations were run with a combination of the following parameters: 
the velocity of the projectile varied between 0.1 and 1.4 a.u., and the 
initial angle of the projectile relative to the $c$ axis of the graphite, $\alpha$, 
varied between $0^{\circ}$ and $90^{\circ}$ as shown in Figure \ref{1b}.
  For simulations of projectiles moving parallel to the graphitic layers, the 
projectiles moved at an angle $\beta$ relative to the $a$ axis, at $0^{\circ}$ 
and $30^{\circ}$ , with checks at $60^{\circ}$ and $90^{\circ}$ as shown in  
Figure \ref{2a}, and moving at distances of $\frac{1}{2}, \frac{1}{4}, 
\frac{1}{8} $and$\frac{1}{16} $ of the spacing between the layers 
from the closest graphitic layer, as shown in Figure \ref{2b}.

  Concerning the charge state of the projectile, previous TDDFT work on 
electronic stopping has been carried out with both ions and atoms 
\cite{Puska, krasheninnikov1, ullah, ahsanzeb}.
  In this type of simulation, the use of a proton or a H atom only changes 
the simulation by one electron in the supercell (out of 129).
  The extent to which the proton drags an electron in its wake is defined 
dynamically, and the established stationary state is independent of the 
initial charge state of the projectile.
  After the initial transient, essentially the same state evolves regardless of 
whether it was initially H$^{+}$ or H, in comparison with other methods in 
which the charge state is defined by hand\cite{ECHENIQUE1990229}. 
  The calculations presented here had 129 electrons in the simulation box,
thereby defining an overall neutral system. The calculations were spin-polarised due to the odd number of electrons in the system.

\section{Results and Discussion}
\label{res}

\subsection{Validation}

\begin{figure}[h!]
\begin{center}
\includegraphics[width=0.43\textwidth]{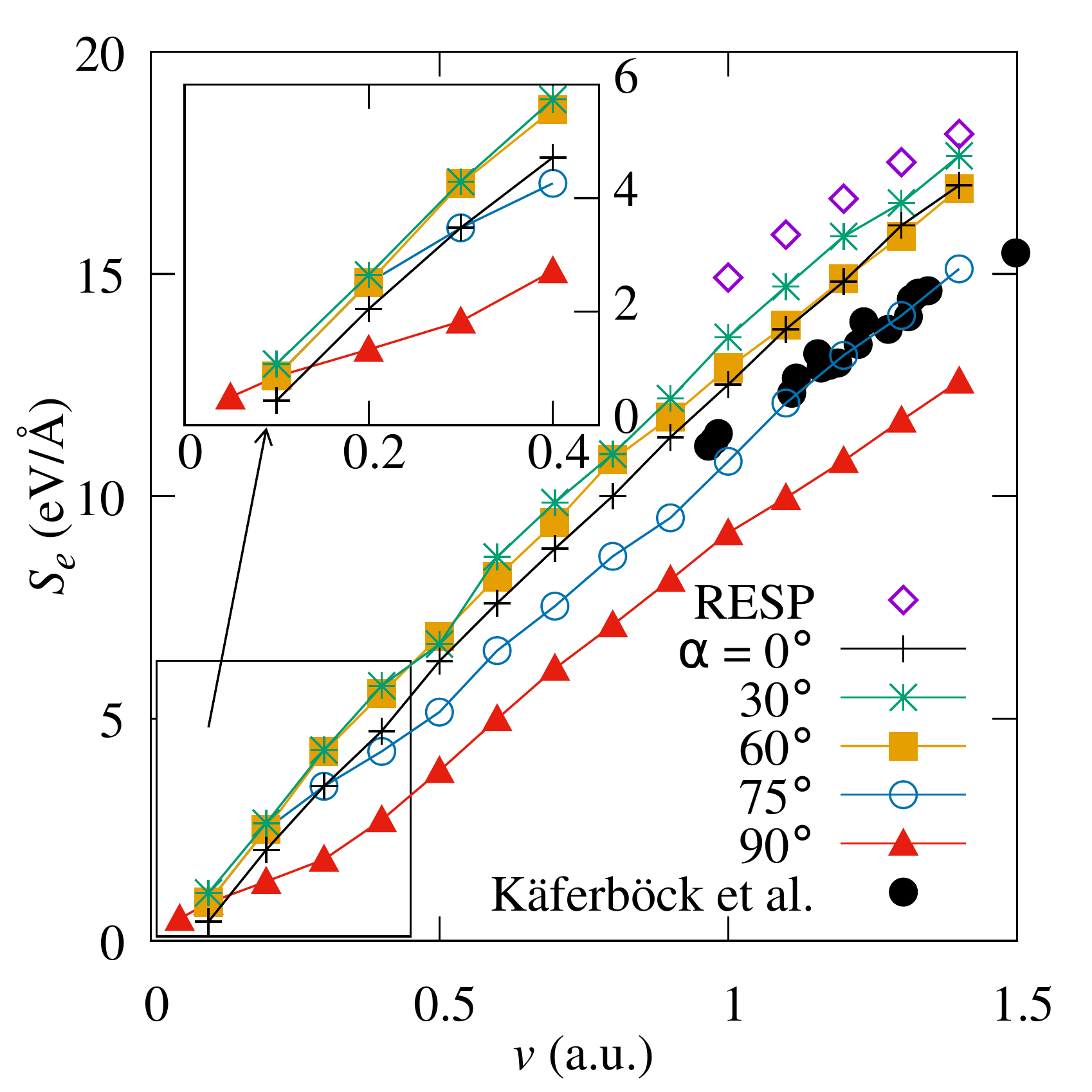}
\end{center}
\caption{Electronic stopping power for a proton shooting through graphite 
at different angles relative to the graphite $c$ axis.$\bullet$ depicts the 
experimental data from the work by K\"aferb\"ock $et\ al.$ \cite{kaferbock}.
  The uncertainty in each data point is $\pm$ 0.6 eV/\AA. RESP is the random electronic stopping power for $\alpha = 90^{\circ}$, averaged over all impact parameters simulated for comparison with Ref.~\onlinecite{PhysRevB.93.035128}.}
\label{fig3red}
\end{figure}

  In order to validate the simulations, the results are compared 
in Figure \ref{fig3red} with experimental data from work by 
K\"aferb\"ock $et\ al.$ \cite{kaferbock}.
  In the velocity range covered by the experimental data, electronic 
stopping dominates the overall stopping power, so the simulations of 
electronic stopping are directly comparable to the K\"aferb\"ock data.
  The Rutherford backscattering experiment used protons with an energy 
of between 20 and 80 keV, corresponding to projectile velocities 
between 1 and 1.7 a.u., with a target of highly ordered pyrolytic graphite, 
although they did not provide angle resolution for the measured $S_{e}$. 
  They consider ion trajectories in all directions. 

  In Figure \ref{fig3red}, the agreement between experiment and theory 
is clear, with the experimental observations lying within the simulation 
range defined by different trajectories.
  The experimental stopping power is closest to that of the higher angle 
simulations, $\alpha = 60^{\circ}$ - $75^{\circ}$.
  As channelling directions were avoided in the experiments, it is likely 
that trajectories close to $\alpha = 90^{\circ}$ contribute little to the 
experimental averaging.
  See a similar consideration in the work of Schleife $et\ al.$ \cite{schleife} 
for a proton moving in aluminium.
  In order to compare the simulations to the experimental data in more 
detail, a model of the distribution of projectile trajectories would be 
needed to calculate an average $S_{e}$ for a particular velocity, which 
involves non-trivial assumptions on the actual trajectories in experimental 
settings. 
  $S_{e}$ gradually diminishes toward the minimum at $\alpha = 90^{\circ}$, 
starting at around $\alpha = 50-60^{\circ}$ where a slight maximum appears, 
especially at low velocities.
 
\subsection{Dependence on $\alpha$}

\begin{figure}[h!]
\includegraphics[width=0.5\textwidth]{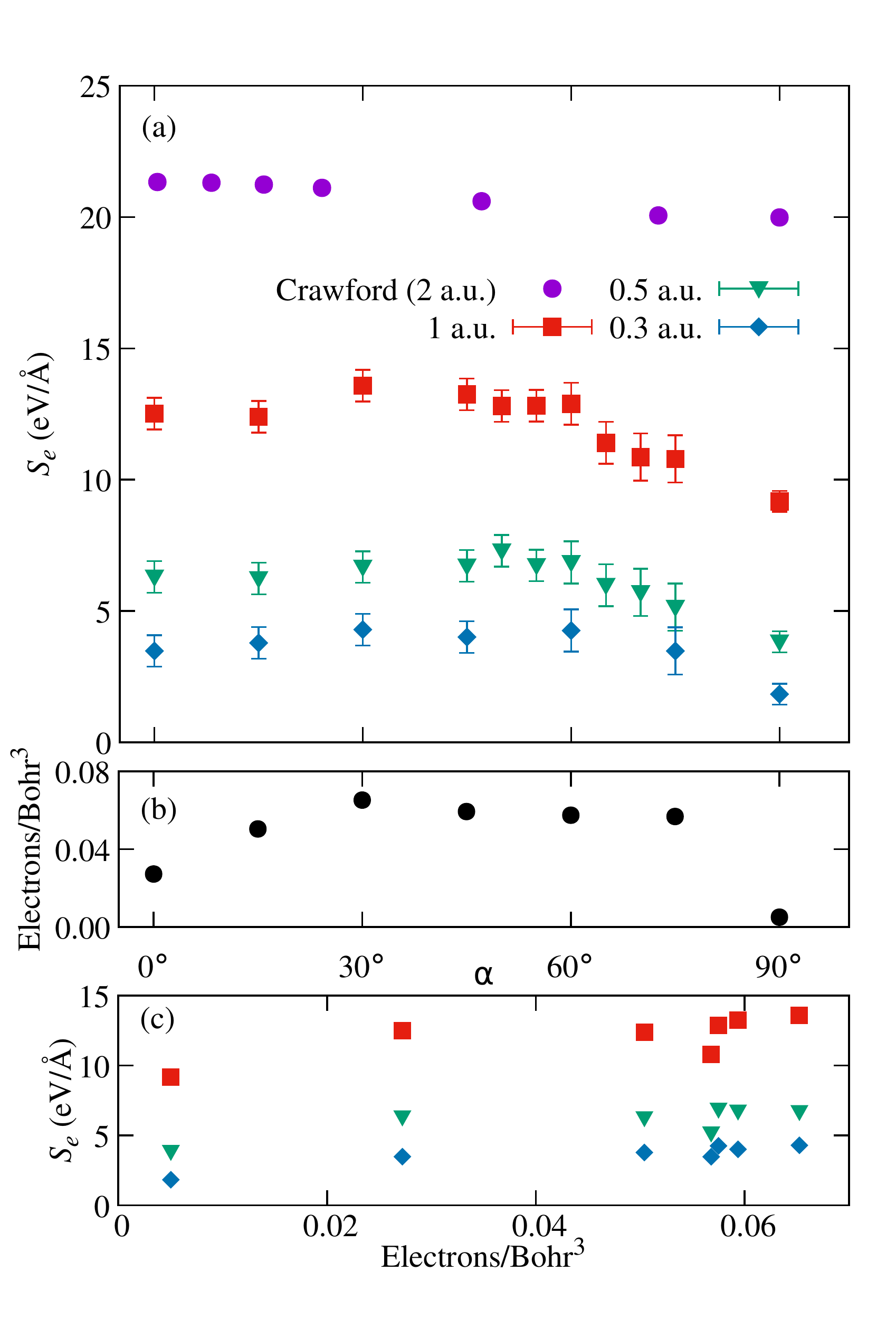}
\caption{(a) Electronic stopping power of a hydrogen atom in graphite 
moving at angle $\alpha$ from the graphite $c$ axis at a velocity of 
0.3, 0.5 and 1.0 a.u..
  The data for $90^{\circ}$ were calculated from simulations with the 
projectile moving midway between two planes of atoms along 
$\beta = 0^{\circ}$ (See Figures \ref{1a} and \ref{2a}).
  The error bars are due to the uncertainty in fitting to the slope of the 
energy plot. 
  (b) Average electron density along the trajectory of the projectile for 
different $\alpha$ values.
  (c) Correlation between local electron density and $S_{e}$}
\label{fig4} 
\end{figure}

  Figure \ref{fig3red} shows the expected overall linear dependence 
of $S_{e}$ in the displayed velocity range, with a slow downward 
bending as velocity increases towards the $S_{e}$ maximum related 
to the Bragg peak, which in graphite is at $\sim$1.9 a.u.
\cite{doi:10.1118/1.597176}.
  The curve for $\alpha = 90^{\circ}$ is clearly different, however, 
corresponding to trajectories parallel to graphitic planes.
  As Figure \ref{fig4} shows more clearly, the electronic stopping power 
decreases  significantly at all velocities between $\alpha = 60^{\circ}$ 
and $\alpha = 90^{\circ}$.
  The data for $\alpha = 90^{\circ}$ correspond to the projectile moving 
midway between two planes of atoms along $\beta = 0^{\circ}$
(See Figures \ref{1a} and \ref{2a}).

  Figure \ref{fig4} also includes the linear-response results of Crawford 
\cite{crawford2} for comparison.
  The lowest velocity considered in that work is $v = 2.0$ a.u., higher 
than those obtained in this work, which accounts for the higher overall 
$S_{e}$.
  They also show a smaller angle dependence, of approximately 10\% 
between a projectile moving along $\alpha = 0^{\circ}$ and $90^{\circ}$ 
at 2 a.u., with smaller differences at higher projectile velocities; significantly 
lower than the equivalent difference for $v = 1.0$ a.u. in our case.
  This is consistent with a further insensitivity with direction at high 
velocities \cite{crawford2}.
Figure 9 of Shukri, Bruneval and Reining's paper \cite{PhysRevB.93.035128} also shows a small difference of up to 3\% in $S_{e}$ between calculations with a projectile moving along $\alpha = 0^{\circ}$ and $90^{\circ}$ in graphite at velocities between 0 and 4 a.u.. As discussed below, that work calculated the random electronic stopping power, which is averaged over all impact parameters, and so is not directly equivalent to the results in this paper.

\subsubsection{Correlation of $S_e$ and electron density}

  Figure \ref{fig4}b shows the average electron density along a given
trajectory versus $\alpha$ for comparison with $S_e(\alpha)$, with the correlation between the two plotted in Figure \ref{fig4}c. 
  The relationship between the electron density and $S_{e}$ is especially 
clear for the low velocities, with both $S_{e}$ and electron density increasing 
from $\alpha = 0^{\circ}$ to $30^{\circ}$, and the lowest $S_{e}$ at 
$90^{\circ}$ corresponding to the lowest electron density. 

  Under the scattering theory formalism developed by Echenique 
and others for jellium \cite{PhysRevB.58.2357, PhysRevA.56.4795, 
PhysRevB.67.245401, echenique86}, the target electron density is 
space independent, a number $n$, and the stopping power is a function 
$S_{e}(v,n)$, which starts at zero for $n=0$ and increases with higher density.
  This results has been generalised to non-homogeneous electron systems, 
with the observation that the stopping power is larger when the 
projectiles traverse regions of higher density.
  This has been seen in previous work in various materials 
\cite{ullah, PhysRevB.58.2357, PhysRevA.56.4795, PhysRevB.67.245401}, 
and is used as a basic assumption in different contexts, as a local-density 
stopping approximation \cite{PhysRevB.67.245401}.
  The results of the simulations presented here are consistent with this; 
between the graphitic layers the electron density is much lower than 
across the layers, and the higher the angle of the projectile relative to 
the $c$ axis the more time it spends in the lower electron density region 
between the layers.
  As a result the projectile interacts less with the electrons of the target 
and so the $S_{e}$ is lower, as seen in Figure \ref{fig4}, obtaining the 
minimum $S_{e}$ for $\alpha = 90^{\circ}$ and the projectile moving 
midway between the planes, that is, the trajectory furthest from the 
carbon atoms and thus with the lowest electron density.  
  It must be noted, however, that although the correlation is clear,
it is not strict, as can be observed for $\alpha=75^{\circ}$, although because of the orientation within the cell, the sampled trajectory both for the evaluation of $S_{e}$ and for the sampling of density values is poorer, which could be behind the larger deviation.

\subsubsection{Channelling}

  The main channelling direction in graphite is along the $c$ axis ($\alpha = 0^{\circ}$), 
but the effect on $S_{e}$ is limited. 
  Only a small depression can be observed for $S_{e}$ for $\alpha=0^{\circ}$ 
as compared with $15^{\circ}$ at low velocity.
  In Figure \ref{fig4} it is visible for $v = 0.3 - 0.5$ a.u., but it is clearly a 
much smaller effect than the one for $\alpha = 90^{\circ}$ (parallel to graphitic layers) even if using perfect 
channelling trajectories, as trajectory 1 in Figure \ref{1a}. 
  These results are consistent with the previous discussion, since the 
average density along that path does not reduce as much as for those 
parallel to and midway between graphitic planes.   

  Channelling in a crystal occurs when a projectile arrives into 
a channel in a trajectory within a small angle from the channel in 
a major crystal direction, and then moves along it undergoing 
small angle scattering, thereby moving along the channel.
  It is customarily expected that, as a result of the lack of nuclear collisions with the target material (beyond the small deflections implied by the channelling 
itself) and thereby reduced total energy loss of the projectile, the 
projectile travels further compared with a random direction 
in the crystal \cite{lindhard1965}.
For light projectiles and velocities above $\sim 0.1$ a.u., however, that effect becomes less important than the fact that $S_{e}$ itself is  expected to be lower along a channelling direction, due to the lower average electron density in a channel;  Schleife, Kanai and Correa \cite{schleife} carried out TDDFT 
simulations of H in Al, comparing projectiles moving along 
channels with off-channelling directions, and found lower $S_{e}$ 
for a projectile moving along a channelling direction than along a 
random non-channelling direction. 
  Channelling in graphite has only been experimentally observed 
when the projectile is moving along the c axis of graphite, 
perpendicular to the graphitic layers \cite{doi:10.1063/1.334210, doi:10.1080/00337577508239479,SCHROYEN1986341}.
  That work used polycrystalline highly oriented pyrolytic 
graphite (HOPG), containing grain boundaries perpendicular 
to the graphitic layers, which would disrupt channelling between 
the layers. In HOPG, the basal planes are closely aligned, but 
alignment along the other axes is difficult to achieve.
  In theory, channelling would also be expected for a projectile 
travelling parallel to the graphitic layers,  and the projectiles moving 
along $\alpha = 90^{\circ}$ do show significantly lower $S_{e}$.  

\begin{figure}[h!]
\includegraphics[width=0.43\textwidth]{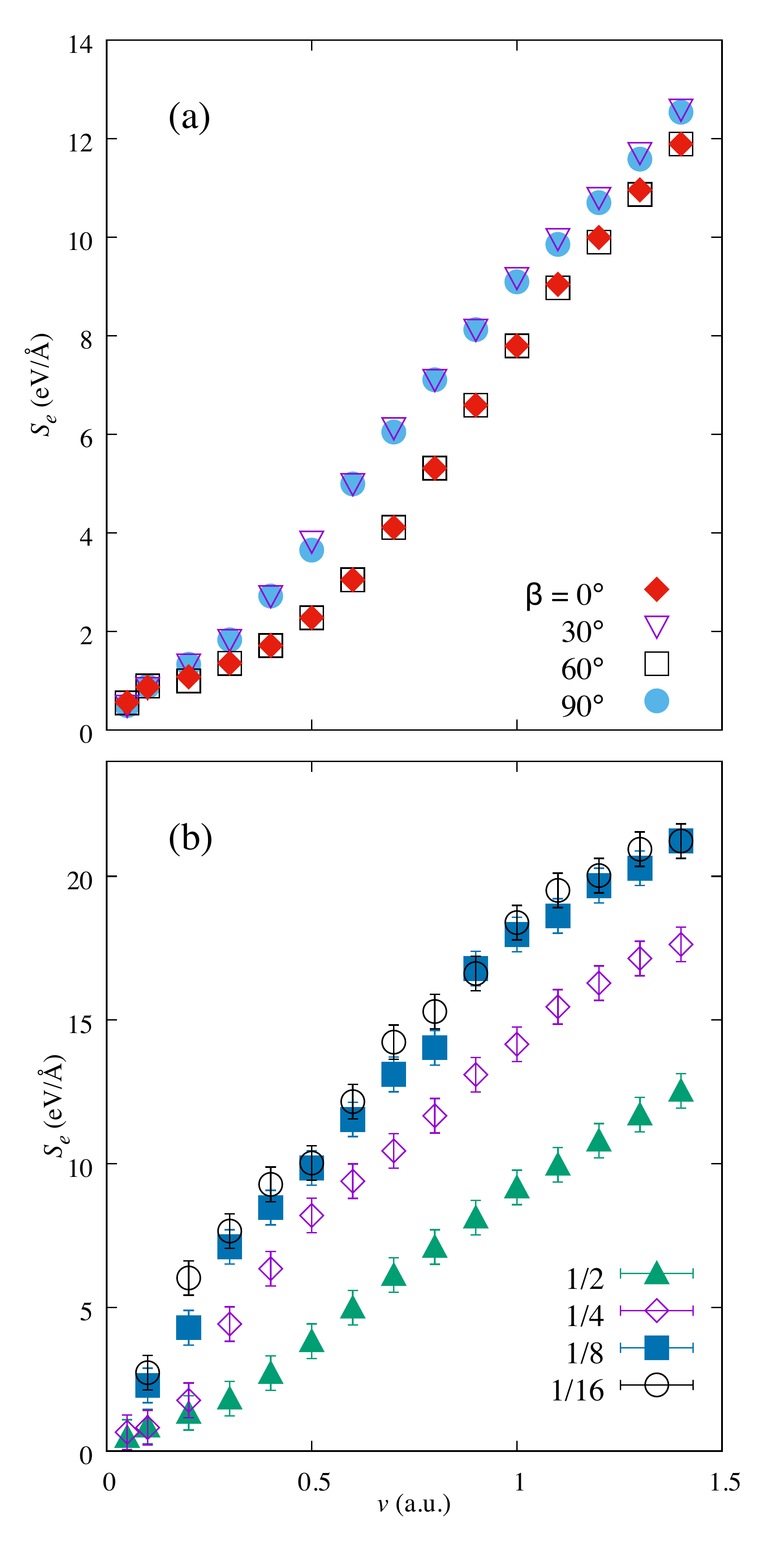}
\caption{(a) shows the electronic stopping power for a projectile moving parallel 
to the graphitic layers at angles $\beta$ from the $a$ axis of graphite, 
midway between the graphitic layers.
  (b) compares the electronic stopping power for a proton moving at different 
distances from the graphitic layers as a fraction of the interplanar distance 
(Figure \ref{2a}).
  The electronic stopping power increases the closer the path of the projectile 
is to a graphitic layer, likely due to the higher electron density closer to the 
planes. }
\label{fig5}
\end{figure}

\subsubsection{Low velocity}

  The behavior of $S_{e}$ in the low velocity end of Figure \ref{fig3red} is 
remarkable.
  On the one hand, the simulations with the projectile moving at angles other 
than $90^{\circ}$ to the $c$ axis appear to show a threshold velocity of 
0.02-0.06 eV below which, extrapolating the calculated data, $S_{e}$ 
appears to be either zero or very small on that scale.
  This is consistent with behavior seen in insulators \cite{ullah} where 
the band gap results in a velocity threshold for $S_{e}$.
  Graphite is effectively a semiconductor in the direction perpendicular 
to the graphitic layers, and, in that sense, this behavior would appear to 
be consistent with what is expected, at least qualitatively. 
  In contrast, the obtained $S_{e}$ shown in 
Figure \ref{fig3red} for $\alpha = 90^{\circ}$, corresponding to a projectile 
moving midway between the graphitic layers, displays a very different 
behavior, with no apparent threshold but rather $S_e \propto v$, but with a 
clear change of slope at $v\sim 0.3$ a.u. displayed by the lowest $S_{e}(v)$ graph 
in Figure \ref{fig3red}.

\subsection{Protons traveling between graphitic planes}

  Figure \ref{fig5} shows the behavior for trajectories parallel to the graphitic 
planes in more detail, with the $S_{e}(v)$ dependence for different orientations 
(Fig. 5a) and different impact parameters (proximity of the trajectory to the 
closest plane, Fig. 5b).
  Starting with Figure 5a, as discussed above, the $S_{e}$ is much lower 
at all velocities and all angles where the projectile is traveling parallel to 
the graphitic layers, as a result of the lower electron density between the layers.
  Due to the hexagonal symmetry of graphite, the trajectories $\beta = 0^{\circ}$ 
and $\beta = 60^{\circ}$ are crystallographically identical. $S_{e}(\beta)$ 
should therefore be periodic with a period of $60^{\circ}$.
  It is expected to be symmetric around $0^{\circ}$ and $ 30^{\circ}$, the 
values of $S_{e}$  for those $\beta$'s representing likely bounds for 
$S_{e}(\beta)$.
  Figure 5a shows $S_{e}$ at 0 and $30^{\circ}$ as a function of velocity.
  The periodicity has been checked with the inclusion of results for 
$\beta = 60^{\circ}$ and $90^{\circ}$. 

   Figure 5(a) shows that the change of slope remains apparent for trajectories 
equidistant from two graphitic planes, irrespective of the $\beta$ angle, 
although for $\beta = 0^{\circ}$ it happens at a slightly larger value of 
$v$ ($v_{\mathrm{K}} \sim 0.4$ a.u.) than for $\beta = 30^{\circ}$ 
($v_{\mathrm{M}}\sim 0.3$ a.u.).
  The former corresponds to the direction of the K-point in reciprocal space, 
while the latter to the M-point direction. Both values are close to the Fermi 
velocity of electrons around the Dirac cone ($v_{F} = 0.37$ a.u.), indicating 
that the change of slope is due to the onset of intra-cone electron-hole 
transitions contributing to the stopping.
  We base this observation on the fact that the electron-hole 
excitations generated by the moving projectile should respect the
relation\cite{ullah}
\begin{equation*}
\mathbf{v} \cdot \Delta \mathbf{k} = \Delta \epsilon
\end{equation*}
being $\Delta \mathbf{k}$ and $\Delta \epsilon$ the momentum and energy
change of the electron, respectively, in the excitation, and $\mathbf{v}$ the
projectile's velocity.
  For $v < v_F$, excitations can only be connecting across cones, while for $v\ge v_F$
the intra-cone channel is open. 

  A similar increase in the $S_{e}$ gradient at velocities between 
0.3 and 0.5 a.u. has been seen in experiments for various systems: 
protons in Au \cite{PhysRevA.75.010901, PhysRevB.78.195122}, 
He in Al \cite{PhysRevLett.107.163201}, and protons and He in 
Cu \cite{PhysRevB.80.205105}, to name a few.
  The change in gradient for the Cu and Au experiments is suggested 
to be a result of interactions with the target's 3$d$ and 5$d$ electrons 
in Cu and Au respectively at higher projectile velocities, where a 
minimum energy transfer is required for the excitation of $d$ electrons 
in both metals \cite{PhysRevA.75.010901}. 
  For He in Al, the slope change is thought to be due to charge-exchange 
processes between the target atoms and projectile \cite{PhysRevLett.107.163201}.
  This again suggests that the increase in gradient is due to additional 
energy loss mechanisms becoming accessible beyond a certain velocity, 
and which, in this case would correspond to the mentioned intra-cone transitions, 
meaning electron-hole-pair formation within the same band and small 
momentum transfer within the Brillouin zone, as the velocity approaches 
the Fermi velocity of the host.

\subsubsection{Impact parameter dependence}

  The impact parameter dependence is shown in Figure \ref{fig5}b.
  It compares the $S_{e}$ for a projectile moving midway between 
the graphitic layers, and at positions $\frac{1}{4}$, $\frac{1}{8}$ and $\frac{1}{16}$ of the 
interplanar distance from a graphitic layer, as shown in Fig.~2(b).
   The $S_{e}$ is higher at all velocities above 0.1 a.u. for the simulations 
 closer to the graphite atoms, corresponding to the higher electron 
 density closer to the graphitic layer. 
  The gradient of the $S_{e}$ plot changes as the velocity increases, 
with a linear region between ~0.5 and 1 a.u., and a slight decrease in 
gradient at higher velocities for both paths as the $S_{e}$ approaches 
a maximum. 
   When the trajectories get closer to either atomic plane (Fig. 5b) 
$S_{e}$ significantly increases as compared to the mid-plane trajectory,
and the clean two-slope structure of Figure \ref{fig5}(a)
is lost, which should be attributed to scattering amplitude effects.

  Trajectories perpendicular to the graphitic planes do not display 
significant impact-parameter dependence, however, unlike what is 
seen for trajectories parallel to the planes. 
  $S_{e}$ increased only by 0.68 eV/\AA\ when changing from 
trajectory 1 in Figure \ref{1a} to trajectory 4 at $v = 0.5$ a.u..

The results of Shukri, Bruneval and Reining \cite{PhysRevB.93.035128} investigated random electronic stopping power, defined as the $S_{e}$ averaged over all impact parameters.  For the in-plane simulations, this is equivalent to averaging $S_{e}$ for all the trajectories at different distances from the graphitic layers. As Figure \ref{fig5} shows, there is a significant increase in $S_{e}$ as the trajectory gets closer to a graphitic layer. The 3\% difference between in-plane and out-of-plane simulations in $S_{e}$ seen by Shukri $et\ al.$ \cite{PhysRevB.93.035128} is therefore consistent with the results in this work. Figure \ref{fig3red} shows the RESP from this work calculated as an average of the four impact parameters simulated for $\alpha = 90^{\circ}$. This simple average oversamples trajectories  close to the  graphitic layers (higher $S_{e}$ and therefore is an overestimate of the RESP; at high velocities we would expect all trajectories to be sampled equally..

\section{Conclusions}
\label{concl}

  Simulations of a hydrogen projectile traveling through graphite 
successfully reproduced experimental results, and provided new 
insights into the effect of the anisotropy of the graphite structure on 
electronic stopping power.
  The electronic stopping power is dependent on the direction of the 
projectile both relative to the graphitic layer normal and parallel to the 
layers.
  Although a clear correlation is found between the local electron 
density traversed by the trajectory in general, at low velocity $S_{e}$ 
displays varied behaviors depending on the direction and impact 
parameter.
  For channelling between planes and low density, a linear $S_{e}$ is 
observed, consistent with (semi)metallic electron conduction, but which 
changes slope when the projectile velocity reaches the Fermi velocity 
of the target.
  For trajectories with dominant component perpendicular to the 
graphitic plane, a threshold is observed at $v \sim 0.05$ a.u., 
consistent with poor electron conduction between planes. 

This work investigates the initial stages of radiation damage; for a fuller understanding of the processes that lead to the final observed radiation damage, simulations must be carried out at longer time scales and with larger simulation sizes. Progression from $S_{e}$ calculations can be envisaged by following the diffusion of the excess electronic energy, and its thermalisation to the ionic motion. In addition to the technical challenges from increased simulation size, theoretical challenges also exist, as Ehrenfest dynamics are known to be inadequate for the simulation of this thermalisation. Beyond Ehrenfest approximations are far more computationally demanding, and therefore present a significant challenge but would offer valuable insights into the progress of stopping processes and the mechanisms of radiation damage.
 
\begin{figure}
\includegraphics[width=0.45\textwidth]{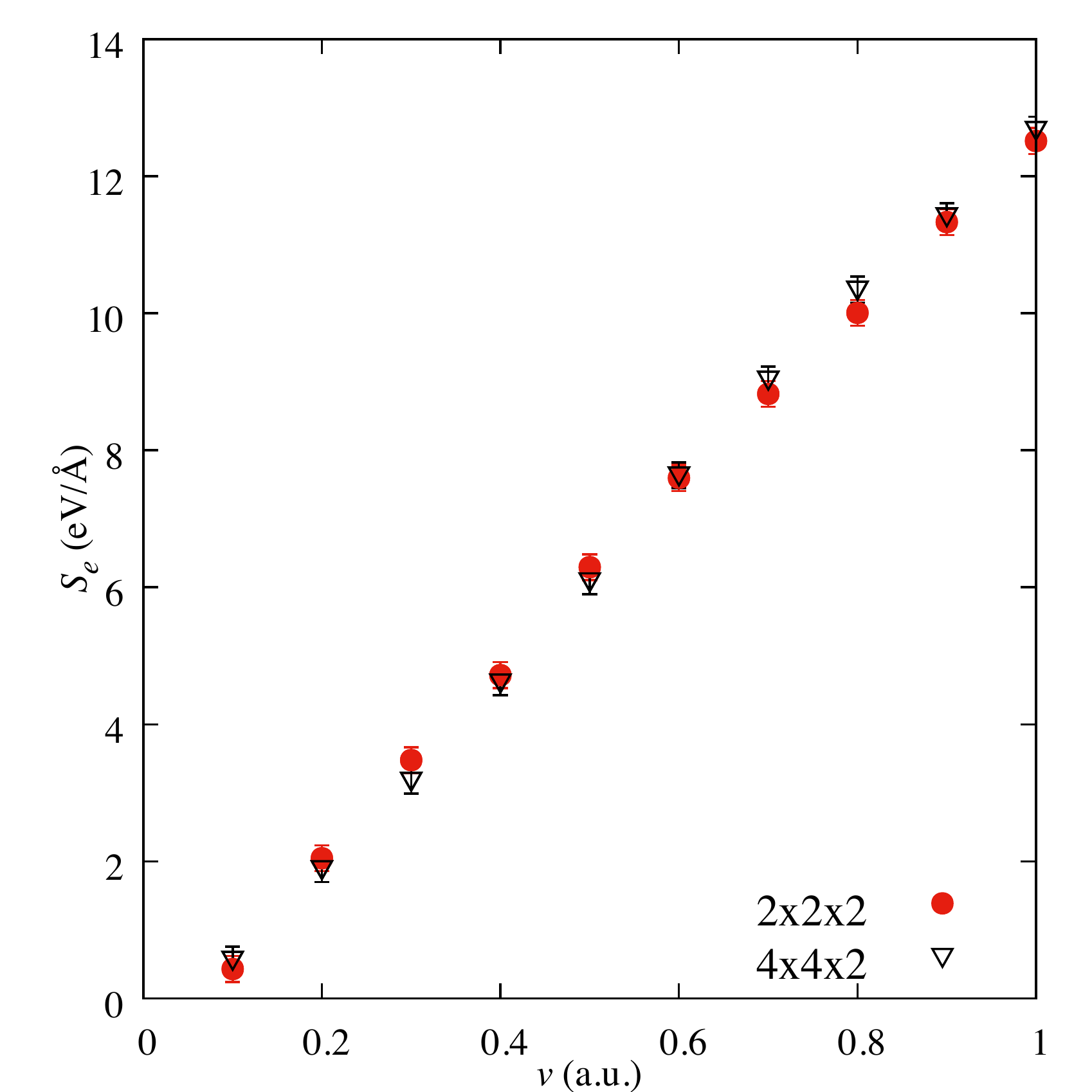}
\caption{Comparison of electronic stopping power in graphite with a 
hydrogen projectile traveling at 0.1 to 1 a.u. with supercell sizes of 
$2\times 2\times 2$ and $4\times 4\times 2$ primitive unit cells.
  In these simulations the projectile moved perpendicular to the graphitic 
planes along the shortest supercell dimension, through the center of a channel.}
\label{cellsize}
\end{figure}

\begin{acknowledgements}

  The authors would like to thank P. Alemany for useful discussions 
on this work.
  J. H. would like to thank R. Ullah for his technical help 
in the initial stages of this work. 
  This work was performed using resources provided by the Cambridge 
Tier-2 system operated by the University of Cambridge Research Computing Service \cite{csd3}, funded by Engineering 
and Physical Sciences Research Council Tier-2  (capital Grant No. EP/P020259/1), 
and DiRAC funding from the Science and Technology Facilities Council 
\cite{dirac}.
  J. H. would like to acknowledge the EPSRC Centre for 
Doctoral Training in Computational Methods for Materials Science for 
funding under Grant No. EP/L015552/1.

\end{acknowledgements}



\section*{Appendix}

  This section describes the testing carried out to generate the initial 
simulation parameters.
  The pseudopotentials of C and H were generated using the scheme 
of Troullier and Martins \cite{troullier} and the corresponding parameters 
are shown in Table \ref{table1}. 

\begin{table}[H]
\caption{Pseudopotential radii for each angular-momentum channel 
of C and H. Lengths are in Bohr.}
\begin{ruledtabular}
\begin{tabular*}{\columnwidth}{@{\extracolsep{\fill}} l|c c c c}
Species & $s$ &$p$ & $d$ &$f$\\
\hline
C (2$s^{2}$2$p^{2}$) & 1.49 & 1.50 & 1.56 & 1.56 \\
\hline
H (1$s^{2}$) & 1.25 & 1.25 & 1.25 & 1.25\\
\end{tabular*}
\end{ruledtabular}
\label{table1}
\end{table}

  Table \ref{table2} gives the parameters needed for the generation of 
the basis set used in this work, following the procedures described in 
Ref.~\citenum{junquera}. The polarisation orbitals were generated by applying an electric field to the orbital according to the procedure implemented in SIESTA and described in Ref.~\onlinecite{siesta}.

\begin{table}[H]
\caption{Cutoff radii $r(\zeta_{1})$ and $r(\zeta_{2})$ in Bohr of the first and second 
$\zeta$ functions of C and H.}
\begin{ruledtabular}
\begin{tabular*}{\columnwidth}{@{\extracolsep{\fill}} l|c c c c}
Species & $n$ &$l$ & $r(\zeta_{1})$ &$r(\zeta_{2})$\\
\hline
C  & 2 & 0 & 4.192 & 3.432 \\
   & 2 & 1 & 4.870 & 3.475 \\
\hline
H  & 1 & 0 & 4.828 & 3.855\\
\end{tabular*}
\end{ruledtabular}
\label{table2}
\end{table}

  A periodic supercell of $2\times 2\times 2$ graphite unit cells was used, 
containing 32 C atoms and a single H atom, with lattice parameters of 
$a$ = 2.461 \AA, $c$ = 6.573 \AA.
  A number of supercell sizes were tested to confirm that the supercell used 
was sufficiently large to give good quality results.
  Figure \ref{cellsize} compares the electronic stopping power for a projectile moving perpendicular to the graphitic layers in $2\times 2\times 2$ and a $4\times 4\times 2$ supercells containing 33 
and 129 atoms.
  There is no significant difference between the electronic stopping powers 
in this velocity range for the two supercell sizes, confirming that the 
$2\times 2\times 2$ supercell is sufficient to produce accurate results. 

  A single k-point ($\Gamma$) was used for the Brillouin zone integrations, 
after testing a ground state calculation with up to 90 k-points for convergence 
and simulation time.
  The difference between $S_{e}$ for one k-point and 96, for trajectory 
number 1 and $v = 1$ a.u., was only 0.2 eV/\AA.
  A timestep of 1 attosecond was used for the low velocity simulations 
up to 1 a.u., and of 0.1 attoseconds for simulations above 1 a.u. 
after testing for the stability of the Crank-Nicolson integrator algorithm 
and the energy change in a TDDFT simulation.
  The plane-wave energy cutoff for real space integration was tested 
between 50 and 400 Ry; the total energy converged at around 150 Ry, 
and a 200 Ry cutoff energy was finally used in the simulations.

\bibliography{articlebib}

\begin{thebibliography}{63}%
\makeatletter
\providecommand \@ifxundefined [1]{%
 \@ifx{#1\undefined}
}%
\providecommand \@ifnum [1]{%
 \ifnum #1\expandafter \@firstoftwo
 \else \expandafter \@secondoftwo
 \fi
}%
\providecommand \@ifx [1]{%
 \ifx #1\expandafter \@firstoftwo
 \else \expandafter \@secondoftwo
 \fi
}%
\providecommand \natexlab [1]{#1}%
\providecommand \enquote  [1]{``#1''}%
\providecommand \bibnamefont  [1]{#1}%
\providecommand \bibfnamefont [1]{#1}%
\providecommand \citenamefont [1]{#1}%
\providecommand \href@noop [0]{\@secondoftwo}%
\providecommand \href [0]{\begingroup \@sanitize@url \@href}%
\providecommand \@href[1]{\@@startlink{#1}\@@href}%
\providecommand \@@href[1]{\endgroup#1\@@endlink}%
\providecommand \@sanitize@url [0]{\catcode `\\12\catcode `\$12\catcode
  `\&12\catcode `\#12\catcode `\^12\catcode `\_12\catcode `\%12\relax}%
\providecommand \@@startlink[1]{}%
\providecommand \@@endlink[0]{}%
\providecommand \url  [0]{\begingroup\@sanitize@url \@url }%
\providecommand \@url [1]{\endgroup\@href {#1}{\urlprefix }}%
\providecommand \urlprefix  [0]{URL }%
\providecommand \Eprint [0]{\href }%
\providecommand \doibase [0]{http://dx.doi.org/}%
\providecommand \selectlanguage [0]{\@gobble}%
\providecommand \bibinfo  [0]{\@secondoftwo}%
\providecommand \bibfield  [0]{\@secondoftwo}%
\providecommand \translation [1]{[#1]}%
\providecommand \BibitemOpen [0]{}%
\providecommand \bibitemStop [0]{}%
\providecommand \bibitemNoStop [0]{.\EOS\space}%
\providecommand \EOS [0]{\spacefactor3000\relax}%
\providecommand \BibitemShut  [1]{\csname bibitem#1\endcsname}%
\let\auto@bib@innerbib\@empty
\bibitem [{\citenamefont {Was}(2007)}]{was}%
  \BibitemOpen
  \bibfield  {author} {\bibinfo {author} {\bibfnamefont {G.~S.}\ \bibnamefont
  {Was}},\ }\href@noop {} {\emph {\bibinfo {title} {Fundamentals of Radiation
  Materials Science}}}\ (\bibinfo  {publisher} {Springer-Verlag},\ \bibinfo
  {year} {2007})\BibitemShut {NoStop}%
\bibitem [{\citenamefont {Begg}\ \emph {et~al.}(2011)\citenamefont {Begg},
  \citenamefont {Stewart},\ and\ \citenamefont {Vens}}]{cancer}%
  \BibitemOpen
  \bibfield  {author} {\bibinfo {author} {\bibfnamefont {A.~C.}\ \bibnamefont
  {Begg}}, \bibinfo {author} {\bibfnamefont {F.~A.}\ \bibnamefont {Stewart}}, \
  and\ \bibinfo {author} {\bibfnamefont {C.}~\bibnamefont {Vens}},\ }\href
  {\doibase 10.1038/nrc3007} {\bibfield  {journal} {\bibinfo  {journal} {Nat
  Rev Cancer}\ }\textbf {\bibinfo {volume} {11}},\ \bibinfo {pages} {239}
  (\bibinfo {year} {2011})}\BibitemShut {NoStop}%
\bibitem [{\citenamefont {Eisen}\ \emph {et~al.}(1972)\citenamefont {Eisen},
  \citenamefont {Clark}, \citenamefont {B⊘ttiger},\ and\ \citenamefont
  {Poate}}]{doi:10.1080/00337577208231165}%
  \BibitemOpen
  \bibfield  {author} {\bibinfo {author} {\bibfnamefont {F.~H.}\ \bibnamefont
  {Eisen}}, \bibinfo {author} {\bibfnamefont {G.~J.}\ \bibnamefont {Clark}},
  \bibinfo {author} {\bibfnamefont {J.}~\bibnamefont {B⊘ttiger}}, \ and\
  \bibinfo {author} {\bibfnamefont {J.~M.}\ \bibnamefont {Poate}},\ }\href
  {\doibase 10.1080/00337577208231165} {\bibfield  {journal} {\bibinfo
  {journal} {Radiation Effects}\ }\textbf {\bibinfo {volume} {13}},\ \bibinfo
  {pages} {93} (\bibinfo {year} {1972})}\BibitemShut {NoStop}%
\bibitem [{\citenamefont {Dygo}\ \emph {et~al.}(1994)\citenamefont {Dygo},
  \citenamefont {Boshart}, \citenamefont {Seiberling},\ and\ \citenamefont
  {Kabachnik}}]{PhysRevA.50.4979}%
  \BibitemOpen
  \bibfield  {author} {\bibinfo {author} {\bibfnamefont {A.}~\bibnamefont
  {Dygo}}, \bibinfo {author} {\bibfnamefont {M.~A.}\ \bibnamefont {Boshart}},
  \bibinfo {author} {\bibfnamefont {L.~E.}\ \bibnamefont {Seiberling}}, \ and\
  \bibinfo {author} {\bibfnamefont {N.~M.}\ \bibnamefont {Kabachnik}},\ }\href
  {\doibase 10.1103/PhysRevA.50.4979} {\bibfield  {journal} {\bibinfo
  {journal} {Phys. Rev. A}\ }\textbf {\bibinfo {volume} {50}},\ \bibinfo
  {pages} {4979} (\bibinfo {year} {1994})}\BibitemShut {NoStop}%
\bibitem [{\citenamefont {dos Santos}\ \emph {et~al.}(1997)\citenamefont {dos
  Santos}, \citenamefont {Grande}, \citenamefont {Behar}, \citenamefont
  {Boudinov},\ and\ \citenamefont {Schiwietz}}]{PhysRevB.55.4332}%
  \BibitemOpen
  \bibfield  {author} {\bibinfo {author} {\bibfnamefont {J.~H.~R.}\
  \bibnamefont {dos Santos}}, \bibinfo {author} {\bibfnamefont {P.~L.}\
  \bibnamefont {Grande}}, \bibinfo {author} {\bibfnamefont {M.}~\bibnamefont
  {Behar}}, \bibinfo {author} {\bibfnamefont {H.}~\bibnamefont {Boudinov}}, \
  and\ \bibinfo {author} {\bibfnamefont {G.}~\bibnamefont {Schiwietz}},\ }\href
  {\doibase 10.1103/PhysRevB.55.4332} {\bibfield  {journal} {\bibinfo
  {journal} {Phys. Rev. B}\ }\textbf {\bibinfo {volume} {55}},\ \bibinfo
  {pages} {4332} (\bibinfo {year} {1997})}\BibitemShut {NoStop}%
\bibitem [{\citenamefont {K\"aferb\"ock}\ \emph {et~al.}(1997)\citenamefont
  {K\"aferb\"ock}, \citenamefont {R\"ossler}, \citenamefont {Necas},
  \citenamefont {Bauer}, \citenamefont {Pe\~nalba}, \citenamefont {Zarate},\
  and\ \citenamefont {Arnau}}]{kaferbock}%
  \BibitemOpen
  \bibfield  {author} {\bibinfo {author} {\bibfnamefont {W.}~\bibnamefont
  {K\"aferb\"ock}}, \bibinfo {author} {\bibfnamefont {W.}~\bibnamefont
  {R\"ossler}}, \bibinfo {author} {\bibfnamefont {V.}~\bibnamefont {Necas}},
  \bibinfo {author} {\bibfnamefont {P.}~\bibnamefont {Bauer}}, \bibinfo
  {author} {\bibfnamefont {M.}~\bibnamefont {Pe\~nalba}}, \bibinfo {author}
  {\bibfnamefont {E.}~\bibnamefont {Zarate}}, \ and\ \bibinfo {author}
  {\bibfnamefont {A.}~\bibnamefont {Arnau}},\ }\href {\doibase
  10.1103/PhysRevB.55.13275} {\bibfield  {journal} {\bibinfo  {journal} {Phys.
  Rev. B}\ }\textbf {\bibinfo {volume} {55}},\ \bibinfo {pages} {13275}
  (\bibinfo {year} {1997})}\BibitemShut {NoStop}%
\bibitem [{\citenamefont {Softky}(1961)}]{softky}%
  \BibitemOpen
  \bibfield  {author} {\bibinfo {author} {\bibfnamefont {S.~D.}\ \bibnamefont
  {Softky}},\ }\href {\doibase 10.1103/PhysRev.123.1685} {\bibfield  {journal}
  {\bibinfo  {journal} {Phys. Rev.}\ }\textbf {\bibinfo {volume} {123}},\
  \bibinfo {pages} {1685} (\bibinfo {year} {1961})}\BibitemShut {NoStop}%
\bibitem [{\citenamefont {Seltzer}\ and\ \citenamefont
  {Berger}(1982)}]{seltzer}%
  \BibitemOpen
  \bibfield  {author} {\bibinfo {author} {\bibfnamefont {S.~M.}\ \bibnamefont
  {Seltzer}}\ and\ \bibinfo {author} {\bibfnamefont {M.~J.}\ \bibnamefont
  {Berger}},\ }\href@noop {} {\bibfield  {journal} {\bibinfo  {journal} {The
  International Journal of Applied Radiation and Isotopes}\ }\textbf {\bibinfo
  {volume} {33}},\ \bibinfo {pages} {1189 } (\bibinfo {year}
  {1982})}\BibitemShut {NoStop}%
\bibitem [{\citenamefont {Crawford}(1990)}]{crawford2}%
  \BibitemOpen
  \bibfield  {author} {\bibinfo {author} {\bibfnamefont {O.~H.}\ \bibnamefont
  {Crawford}},\ }\href {\doibase 10.1103/PhysRevA.42.1390} {\bibfield
  {journal} {\bibinfo  {journal} {Phys. Rev. A}\ }\textbf {\bibinfo {volume}
  {42}},\ \bibinfo {pages} {1390} (\bibinfo {year} {1990})}\BibitemShut
  {NoStop}%
\bibitem [{\citenamefont {Shukri}\ \emph {et~al.}(2016)\citenamefont {Shukri},
  \citenamefont {Bruneval},\ and\ \citenamefont
  {Reining}}]{PhysRevB.93.035128}%
  \BibitemOpen
  \bibfield  {author} {\bibinfo {author} {\bibfnamefont {A.~A.}\ \bibnamefont
  {Shukri}}, \bibinfo {author} {\bibfnamefont {F.}~\bibnamefont {Bruneval}}, \
  and\ \bibinfo {author} {\bibfnamefont {L.}~\bibnamefont {Reining}},\ }\href
  {\doibase 10.1103/PhysRevB.93.035128} {\bibfield  {journal} {\bibinfo
  {journal} {Phys. Rev. B}\ }\textbf {\bibinfo {volume} {93}},\ \bibinfo
  {pages} {035128} (\bibinfo {year} {2016})}\BibitemShut {NoStop}%
\bibitem [{\citenamefont {Echenique}\ \emph {et~al.}(1981)\citenamefont
  {Echenique}, \citenamefont {Nieminen},\ and\ \citenamefont
  {Ritchie}}]{echenique81}%
  \BibitemOpen
  \bibfield  {author} {\bibinfo {author} {\bibfnamefont {P.}~\bibnamefont
  {Echenique}}, \bibinfo {author} {\bibfnamefont {R.}~\bibnamefont {Nieminen}},
  \ and\ \bibinfo {author} {\bibfnamefont {R.}~\bibnamefont {Ritchie}},\
  }\href@noop {} {\bibfield  {journal} {\bibinfo  {journal} {Solid State
  Communications}\ }\textbf {\bibinfo {volume} {37}},\ \bibinfo {pages} {779 }
  (\bibinfo {year} {1981})}\BibitemShut {NoStop}%
\bibitem [{\citenamefont {Echenique}\ \emph {et~al.}(1986)\citenamefont
  {Echenique}, \citenamefont {Nieminen}, \citenamefont {Ashley},\ and\
  \citenamefont {Ritchie}}]{echenique86}%
  \BibitemOpen
  \bibfield  {author} {\bibinfo {author} {\bibfnamefont {P.~M.}\ \bibnamefont
  {Echenique}}, \bibinfo {author} {\bibfnamefont {R.~M.}\ \bibnamefont
  {Nieminen}}, \bibinfo {author} {\bibfnamefont {J.~C.}\ \bibnamefont
  {Ashley}}, \ and\ \bibinfo {author} {\bibfnamefont {R.~H.}\ \bibnamefont
  {Ritchie}},\ }\href {\doibase 10.1103/PhysRevA.33.897} {\bibfield  {journal}
  {\bibinfo  {journal} {Phys. Rev. A}\ }\textbf {\bibinfo {volume} {33}},\
  \bibinfo {pages} {897} (\bibinfo {year} {1986})}\BibitemShut {NoStop}%
\bibitem [{\citenamefont {Ferrell}\ and\ \citenamefont
  {Ritchie}(1977)}]{ferrell}%
  \BibitemOpen
  \bibfield  {author} {\bibinfo {author} {\bibfnamefont {T.~L.}\ \bibnamefont
  {Ferrell}}\ and\ \bibinfo {author} {\bibfnamefont {R.~H.}\ \bibnamefont
  {Ritchie}},\ }\href {\doibase 10.1103/PhysRevB.16.115} {\bibfield  {journal}
  {\bibinfo  {journal} {Phys. Rev. B}\ }\textbf {\bibinfo {volume} {16}},\
  \bibinfo {pages} {115} (\bibinfo {year} {1977})}\BibitemShut {NoStop}%
\bibitem [{\citenamefont {Lindhard}(1954)}]{lindhard1954}%
  \BibitemOpen
  \bibfield  {author} {\bibinfo {author} {\bibfnamefont {J.}~\bibnamefont
  {Lindhard}},\ }\href@noop {} {\bibfield  {journal} {\bibinfo  {journal}
  {Matematisk-fysiske Meddelelser}\ }\textbf {\bibinfo {volume} {28}} (\bibinfo
  {year} {1954})}\BibitemShut {NoStop}%
\bibitem [{\citenamefont {Echenique}\ \emph {et~al.}(1990)\citenamefont
  {Echenique}, \citenamefont {Flores},\ and\ \citenamefont
  {Ritchie}}]{ECHENIQUE1990229}%
  \BibitemOpen
  \bibfield  {author} {\bibinfo {author} {\bibfnamefont {P.}~\bibnamefont
  {Echenique}}, \bibinfo {author} {\bibfnamefont {F.}~\bibnamefont {Flores}}, \
  and\ \bibinfo {author} {\bibfnamefont {R.}~\bibnamefont {Ritchie}},\ }\href
  {\doibase https://doi.org/10.1016/S0081-1947(08)60325-2} {\bibfield
  {journal} {\bibinfo  {journal} {Solid State Phys.}\ }\textbf {\bibinfo
  {volume} {43}},\ \bibinfo {pages} {229 } (\bibinfo {year}
  {1990})}\BibitemShut {NoStop}%
\bibitem [{\citenamefont {Sigmund}(2014)}]{sigmund2014}%
  \BibitemOpen
  \bibfield  {author} {\bibinfo {author} {\bibfnamefont {P.}~\bibnamefont
  {Sigmund}},\ }\href {\doibase 10.1007/978-3-319-05564-0} {\emph {\bibinfo
  {title} {Particle Penetration and Radiation Effects Volume 2}}}\ (\bibinfo
  {publisher} {Springer International Publishing},\ \bibinfo {year}
  {2014})\BibitemShut {NoStop}%
\bibitem [{\citenamefont {Mason}\ \emph {et~al.}(2007)\citenamefont {Mason},
  \citenamefont {le~Page}, \citenamefont {Race}, \citenamefont {Foulkes},
  \citenamefont {Finnis},\ and\ \citenamefont {Sutton}}]{mason2007}%
  \BibitemOpen
  \bibfield  {author} {\bibinfo {author} {\bibfnamefont {D.~R.}\ \bibnamefont
  {Mason}}, \bibinfo {author} {\bibfnamefont {J.}~\bibnamefont {le~Page}},
  \bibinfo {author} {\bibfnamefont {C.~P.}\ \bibnamefont {Race}}, \bibinfo
  {author} {\bibfnamefont {W.~M.~C.}\ \bibnamefont {Foulkes}}, \bibinfo
  {author} {\bibfnamefont {M.~W.}\ \bibnamefont {Finnis}}, \ and\ \bibinfo
  {author} {\bibfnamefont {A.~P.}\ \bibnamefont {Sutton}},\ }\href
  {http://stacks.iop.org/0953-8984/19/i=43/a=436209} {\bibfield  {journal}
  {\bibinfo  {journal} {J. Phys. Condens. Matter}\ }\textbf {\bibinfo {volume}
  {19}},\ \bibinfo {pages} {436209} (\bibinfo {year} {2007})}\BibitemShut
  {NoStop}%
\bibitem [{\citenamefont {Race}\ \emph {et~al.}(2010)\citenamefont {Race},
  \citenamefont {Mason}, \citenamefont {Finnis}, \citenamefont {Foulkes},
  \citenamefont {Horsfield},\ and\ \citenamefont {Sutton}}]{race2010}%
  \BibitemOpen
  \bibfield  {author} {\bibinfo {author} {\bibfnamefont {C.~P.}\ \bibnamefont
  {Race}}, \bibinfo {author} {\bibfnamefont {D.~R.}\ \bibnamefont {Mason}},
  \bibinfo {author} {\bibfnamefont {M.~W.}\ \bibnamefont {Finnis}}, \bibinfo
  {author} {\bibfnamefont {W.~M.~C.}\ \bibnamefont {Foulkes}}, \bibinfo
  {author} {\bibfnamefont {A.~P.}\ \bibnamefont {Horsfield}}, \ and\ \bibinfo
  {author} {\bibfnamefont {A.~P.}\ \bibnamefont {Sutton}},\ }\href
  {http://stacks.iop.org/0034-4885/73/i=11/a=116501} {\bibfield  {journal}
  {\bibinfo  {journal} {Rep. Prog. Phys.}\ }\textbf {\bibinfo {volume} {73}},\
  \bibinfo {pages} {116501} (\bibinfo {year} {2010})}\BibitemShut {NoStop}%
\bibitem [{\citenamefont {Krasheninnikov}\ \emph {et~al.}(2007)\citenamefont
  {Krasheninnikov}, \citenamefont {Miyamoto},\ and\ \citenamefont
  {Tom\'anek}}]{krasheninnikov1}%
  \BibitemOpen
  \bibfield  {author} {\bibinfo {author} {\bibfnamefont {A.~V.}\ \bibnamefont
  {Krasheninnikov}}, \bibinfo {author} {\bibfnamefont {Y.}~\bibnamefont
  {Miyamoto}}, \ and\ \bibinfo {author} {\bibfnamefont {D.}~\bibnamefont
  {Tom\'anek}},\ }\href {\doibase 10.1103/PhysRevLett.99.016104} {\bibfield
  {journal} {\bibinfo  {journal} {Phys. Rev. Lett.}\ }\textbf {\bibinfo
  {volume} {99}},\ \bibinfo {pages} {016104} (\bibinfo {year}
  {2007})}\BibitemShut {NoStop}%
\bibitem [{\citenamefont {Pruneda}\ \emph {et~al.}(2007)\citenamefont
  {Pruneda}, \citenamefont {S\'anchez-Portal}, \citenamefont {Arnau},
  \citenamefont {Juaristi},\ and\ \citenamefont {Artacho}}]{pruneda}%
  \BibitemOpen
  \bibfield  {author} {\bibinfo {author} {\bibfnamefont {J.~M.}\ \bibnamefont
  {Pruneda}}, \bibinfo {author} {\bibfnamefont {D.}~\bibnamefont
  {S\'anchez-Portal}}, \bibinfo {author} {\bibfnamefont {A.}~\bibnamefont
  {Arnau}}, \bibinfo {author} {\bibfnamefont {J.~I.}\ \bibnamefont {Juaristi}},
  \ and\ \bibinfo {author} {\bibfnamefont {E.}~\bibnamefont {Artacho}},\ }\href
  {\doibase 10.1103/PhysRevLett.99.235501} {\bibfield  {journal} {\bibinfo
  {journal} {Phys. Rev. Lett.}\ }\textbf {\bibinfo {volume} {99}},\ \bibinfo
  {pages} {235501} (\bibinfo {year} {2007})}\BibitemShut {NoStop}%
\bibitem [{\citenamefont {Zeb}\ \emph {et~al.}(2012)\citenamefont {Zeb},
  \citenamefont {Kohanoff}, \citenamefont {S\'anchez-Portal}, \citenamefont
  {Arnau}, \citenamefont {Juaristi},\ and\ \citenamefont
  {Artacho}}]{ahsanzeb2}%
  \BibitemOpen
  \bibfield  {author} {\bibinfo {author} {\bibfnamefont {M.~A.}\ \bibnamefont
  {Zeb}}, \bibinfo {author} {\bibfnamefont {J.}~\bibnamefont {Kohanoff}},
  \bibinfo {author} {\bibfnamefont {D.}~\bibnamefont {S\'anchez-Portal}},
  \bibinfo {author} {\bibfnamefont {A.}~\bibnamefont {Arnau}}, \bibinfo
  {author} {\bibfnamefont {J.~I.}\ \bibnamefont {Juaristi}}, \ and\ \bibinfo
  {author} {\bibfnamefont {E.}~\bibnamefont {Artacho}},\ }\href {\doibase
  10.1103/PhysRevLett.108.225504} {\bibfield  {journal} {\bibinfo  {journal}
  {Phys. Rev. Lett.}\ }\textbf {\bibinfo {volume} {108}},\ \bibinfo {pages}
  {225504} (\bibinfo {year} {2012})}\BibitemShut {NoStop}%
\bibitem [{\citenamefont {Correa}\ \emph {et~al.}(2012)\citenamefont {Correa},
  \citenamefont {Kohanoff}, \citenamefont {Artacho}, \citenamefont
  {S\'anchez-Portal},\ and\ \citenamefont {Caro}}]{correa}%
  \BibitemOpen
  \bibfield  {author} {\bibinfo {author} {\bibfnamefont {A.~A.}\ \bibnamefont
  {Correa}}, \bibinfo {author} {\bibfnamefont {J.}~\bibnamefont {Kohanoff}},
  \bibinfo {author} {\bibfnamefont {E.}~\bibnamefont {Artacho}}, \bibinfo
  {author} {\bibfnamefont {D.}~\bibnamefont {S\'anchez-Portal}}, \ and\
  \bibinfo {author} {\bibfnamefont {A.}~\bibnamefont {Caro}},\ }\href {\doibase
  10.1103/PhysRevLett.108.213201} {\bibfield  {journal} {\bibinfo  {journal}
  {Phys. Rev. Lett.}\ }\textbf {\bibinfo {volume} {108}},\ \bibinfo {pages}
  {213201} (\bibinfo {year} {2012})}\BibitemShut {NoStop}%
\bibitem [{\citenamefont {Zeb}\ \emph {et~al.}(2013)\citenamefont {Zeb},
  \citenamefont {Kohanoff}, \citenamefont {S\'anchez-Portal},\ and\
  \citenamefont {Artacho}}]{ahsanzeb}%
  \BibitemOpen
  \bibfield  {author} {\bibinfo {author} {\bibfnamefont {M.~A.}\ \bibnamefont
  {Zeb}}, \bibinfo {author} {\bibfnamefont {J.}~\bibnamefont {Kohanoff}},
  \bibinfo {author} {\bibfnamefont {D.}~\bibnamefont {S\'anchez-Portal}}, \
  and\ \bibinfo {author} {\bibfnamefont {E.}~\bibnamefont {Artacho}},\ }\href
  {\doibase http://dx.doi.org/10.1016/j.nimb.2012.12.022} {\bibfield  {journal}
  {\bibinfo  {journal} {Nucl. Instrum. Meth. B}\ }\textbf {\bibinfo {volume}
  {303}},\ \bibinfo {pages} {59 } (\bibinfo {year} {2013})}\BibitemShut
  {NoStop}%
\bibitem [{\citenamefont {Ojanper\"a}\ \emph {et~al.}(2014)\citenamefont
  {Ojanper\"a}, \citenamefont {Krasheninnikov},\ and\ \citenamefont
  {Puska}}]{Puska}%
  \BibitemOpen
  \bibfield  {author} {\bibinfo {author} {\bibfnamefont {A.}~\bibnamefont
  {Ojanper\"a}}, \bibinfo {author} {\bibfnamefont {A.~V.}\ \bibnamefont
  {Krasheninnikov}}, \ and\ \bibinfo {author} {\bibfnamefont {M.}~\bibnamefont
  {Puska}},\ }\href {\doibase 10.1103/PhysRevB.89.035120} {\bibfield  {journal}
  {\bibinfo  {journal} {Phys. Rev. B}\ }\textbf {\bibinfo {volume} {89}},\
  \bibinfo {pages} {035120} (\bibinfo {year} {2014})}\BibitemShut {NoStop}%
\bibitem [{\citenamefont {Caro}\ \emph {et~al.}(2015)\citenamefont {Caro},
  \citenamefont {Correa}, \citenamefont {Tamm}, \citenamefont {Samolyuk},\ and\
  \citenamefont {Stocks}}]{caro2015}%
  \BibitemOpen
  \bibfield  {author} {\bibinfo {author} {\bibfnamefont {A.}~\bibnamefont
  {Caro}}, \bibinfo {author} {\bibfnamefont {A.~A.}\ \bibnamefont {Correa}},
  \bibinfo {author} {\bibfnamefont {A.}~\bibnamefont {Tamm}}, \bibinfo {author}
  {\bibfnamefont {G.~D.}\ \bibnamefont {Samolyuk}}, \ and\ \bibinfo {author}
  {\bibfnamefont {G.~M.}\ \bibnamefont {Stocks}},\ }\href {\doibase
  10.1103/PhysRevB.92.144309} {\bibfield  {journal} {\bibinfo  {journal} {Phys.
  Rev. B}\ }\textbf {\bibinfo {volume} {92}},\ \bibinfo {pages} {144309}
  (\bibinfo {year} {2015})}\BibitemShut {NoStop}%
\bibitem [{\citenamefont {Schleife}\ \emph {et~al.}(2015)\citenamefont
  {Schleife}, \citenamefont {Kanai},\ and\ \citenamefont {Correa}}]{schleife}%
  \BibitemOpen
  \bibfield  {author} {\bibinfo {author} {\bibfnamefont {A.}~\bibnamefont
  {Schleife}}, \bibinfo {author} {\bibfnamefont {Y.}~\bibnamefont {Kanai}}, \
  and\ \bibinfo {author} {\bibfnamefont {A.~A.}\ \bibnamefont {Correa}},\
  }\href {\doibase 10.1103/PhysRevB.91.014306} {\bibfield  {journal} {\bibinfo
  {journal} {Phys. Rev. B}\ }\textbf {\bibinfo {volume} {91}},\ \bibinfo
  {pages} {014306} (\bibinfo {year} {2015})}\BibitemShut {NoStop}%
\bibitem [{\citenamefont {Ullah}\ \emph {et~al.}(2015)\citenamefont {Ullah},
  \citenamefont {Corsetti}, \citenamefont {S\'anchez-Portal},\ and\
  \citenamefont {Artacho}}]{ullah}%
  \BibitemOpen
  \bibfield  {author} {\bibinfo {author} {\bibfnamefont {R.}~\bibnamefont
  {Ullah}}, \bibinfo {author} {\bibfnamefont {F.}~\bibnamefont {Corsetti}},
  \bibinfo {author} {\bibfnamefont {D.}~\bibnamefont {S\'anchez-Portal}}, \
  and\ \bibinfo {author} {\bibfnamefont {E.}~\bibnamefont {Artacho}},\ }\href
  {\doibase 10.1103/PhysRevB.91.125203} {\bibfield  {journal} {\bibinfo
  {journal} {Phys. Rev. B}\ }\textbf {\bibinfo {volume} {91}},\ \bibinfo
  {pages} {125203} (\bibinfo {year} {2015})}\BibitemShut {NoStop}%
\bibitem [{\citenamefont {Caro}\ \emph {et~al.}(2017)\citenamefont {Caro},
  \citenamefont {Correa}, \citenamefont {Artacho},\ and\ \citenamefont
  {Caro}}]{caro2017}%
  \BibitemOpen
  \bibfield  {author} {\bibinfo {author} {\bibfnamefont {M.}~\bibnamefont
  {Caro}}, \bibinfo {author} {\bibfnamefont {A.~A.}\ \bibnamefont {Correa}},
  \bibinfo {author} {\bibfnamefont {E.}~\bibnamefont {Artacho}}, \ and\
  \bibinfo {author} {\bibfnamefont {A.}~\bibnamefont {Caro}},\ }\href {\doibase
  10.1038/s41598-017-02780-3} {\bibfield  {journal} {\bibinfo  {journal} {Sci.
  Rep.}\ }\textbf {\bibinfo {volume} {7}},\ \bibinfo {pages} {2045} (\bibinfo
  {year} {2017})}\BibitemShut {NoStop}%
\bibitem [{\citenamefont {Bubin}\ \emph {et~al.}(2012)\citenamefont {Bubin},
  \citenamefont {Wang}, \citenamefont {Pantelides},\ and\ \citenamefont
  {Varga}}]{PhysRevB.85.235435}%
  \BibitemOpen
  \bibfield  {author} {\bibinfo {author} {\bibfnamefont {S.}~\bibnamefont
  {Bubin}}, \bibinfo {author} {\bibfnamefont {B.}~\bibnamefont {Wang}},
  \bibinfo {author} {\bibfnamefont {S.}~\bibnamefont {Pantelides}}, \ and\
  \bibinfo {author} {\bibfnamefont {K.}~\bibnamefont {Varga}},\ }\href
  {\doibase 10.1103/PhysRevB.85.235435} {\bibfield  {journal} {\bibinfo
  {journal} {Phys. Rev. B}\ }\textbf {\bibinfo {volume} {85}},\ \bibinfo
  {pages} {235435} (\bibinfo {year} {2012})}\BibitemShut {NoStop}%
\bibitem [{\citenamefont {{Lee, Cheng-Wei}}\ and\ \citenamefont {{Schleife,
  Andr\'e}}(2018)}]{refId0}%
  \BibitemOpen
  \bibfield  {author} {\bibinfo {author} {\bibnamefont {{Lee, Cheng-Wei}}}\
  and\ \bibinfo {author} {\bibnamefont {{Schleife, Andr\'e}}},\ }\href
  {\doibase 10.1140/epjb/e2018-90204-8} {\bibfield  {journal} {\bibinfo
  {journal} {Eur. Phys. J. B}\ }\textbf {\bibinfo {volume} {91}},\ \bibinfo
  {pages} {222} (\bibinfo {year} {2018})}\BibitemShut {NoStop}%
\bibitem [{\citenamefont {Yost}\ \emph {et~al.}(2017)\citenamefont {Yost},
  \citenamefont {Yao},\ and\ \citenamefont {Kanai}}]{PhysRevB.96.115134}%
  \BibitemOpen
  \bibfield  {author} {\bibinfo {author} {\bibfnamefont {D.~C.}\ \bibnamefont
  {Yost}}, \bibinfo {author} {\bibfnamefont {Y.}~\bibnamefont {Yao}}, \ and\
  \bibinfo {author} {\bibfnamefont {Y.}~\bibnamefont {Kanai}},\ }\href
  {\doibase 10.1103/PhysRevB.96.115134} {\bibfield  {journal} {\bibinfo
  {journal} {Phys. Rev. B}\ }\textbf {\bibinfo {volume} {96}},\ \bibinfo
  {pages} {115134} (\bibinfo {year} {2017})}\BibitemShut {NoStop}%
\bibitem [{\citenamefont {Chung}(2002)}]{chung2002}%
  \BibitemOpen
  \bibfield  {author} {\bibinfo {author} {\bibfnamefont {D.~D.~L.}\
  \bibnamefont {Chung}},\ }\href {\doibase 10.1023/A:1014915307738} {\bibfield
  {journal} {\bibinfo  {journal} {J. Mater. Sci.}\ }\textbf {\bibinfo {volume}
  {37}},\ \bibinfo {pages} {1475} (\bibinfo {year} {2002})}\BibitemShut
  {NoStop}%
\bibitem [{\citenamefont {Cazaux}(1970)}]{Cazaux1970545}%
  \BibitemOpen
  \bibfield  {author} {\bibinfo {author} {\bibfnamefont {J.}~\bibnamefont
  {Cazaux}},\ }\href
  {https://www.sciencedirect.com/science/article/pii/0038109870903017}
  {\bibfield  {journal} {\bibinfo  {journal} {Solid State Commun.}\ }\textbf
  {\bibinfo {volume} {8}},\ \bibinfo {pages} {545 } (\bibinfo {year}
  {1970})}\BibitemShut {NoStop}%
\bibitem [{\citenamefont {Yagi}\ \emph {et~al.}(2004)\citenamefont {Yagi},
  \citenamefont {Iwata}, \citenamefont {Urai},\ and\ \citenamefont
  {Ogiwara}}]{YAGI20049}%
  \BibitemOpen
  \bibfield  {author} {\bibinfo {author} {\bibfnamefont {E.}~\bibnamefont
  {Yagi}}, \bibinfo {author} {\bibfnamefont {T.}~\bibnamefont {Iwata}},
  \bibinfo {author} {\bibfnamefont {T.}~\bibnamefont {Urai}}, \ and\ \bibinfo
  {author} {\bibfnamefont {K.}~\bibnamefont {Ogiwara}},\ }\href {\doibase
  https://doi.org/10.1016/j.jnucmat.2004.04.321} {\bibfield  {journal}
  {\bibinfo  {journal} {J. of Nucl. Mater.}\ }\textbf {\bibinfo {volume}
  {334}},\ \bibinfo {pages} {9 } (\bibinfo {year} {2004})}\BibitemShut
  {NoStop}%
\bibitem [{\citenamefont {Tsolakidis}\ \emph {et~al.}(2002)\citenamefont
  {Tsolakidis}, \citenamefont {S\'anchez-Portal},\ and\ \citenamefont
  {Martin}}]{tsolakidis}%
  \BibitemOpen
  \bibfield  {author} {\bibinfo {author} {\bibfnamefont {A.}~\bibnamefont
  {Tsolakidis}}, \bibinfo {author} {\bibfnamefont {D.}~\bibnamefont
  {S\'anchez-Portal}}, \ and\ \bibinfo {author} {\bibfnamefont {R.~M.}\
  \bibnamefont {Martin}},\ }\href {\doibase 10.1103/PhysRevB.66.235416}
  {\bibfield  {journal} {\bibinfo  {journal} {Phys. Rev. B}\ }\textbf {\bibinfo
  {volume} {66}},\ \bibinfo {pages} {235416} (\bibinfo {year}
  {2002})}\BibitemShut {NoStop}%
\bibitem [{\citenamefont {Ullah}\ \emph {et~al.}(2019)\citenamefont {Ullah},
  \citenamefont {Garaizar}, \citenamefont {Corsetti}, \citenamefont {Kohanoff},
  \citenamefont {Sanchez-Portal},\ and\ \citenamefont {Artacho}}]{ullah2019}%
  \BibitemOpen
  \bibfield  {author} {\bibinfo {author} {\bibfnamefont {R.}~\bibnamefont
  {Ullah}}, \bibinfo {author} {\bibfnamefont {A.}~\bibnamefont {Garaizar}},
  \bibinfo {author} {\bibfnamefont {F.}~\bibnamefont {Corsetti}}, \bibinfo
  {author} {\bibfnamefont {J.}~\bibnamefont {Kohanoff}}, \bibinfo {author}
  {\bibfnamefont {D.}~\bibnamefont {Sanchez-Portal}}, \ and\ \bibinfo {author}
  {\bibfnamefont {E.}~\bibnamefont {Artacho}},\ }\href@noop {} {\ ,\ \bibinfo
  {pages} {forthcoming} (\bibinfo {year} {2019})}\BibitemShut {NoStop}%
\bibitem [{\citenamefont {Soler}\ \emph {et~al.}(2002)\citenamefont {Soler},
  \citenamefont {Artacho}, \citenamefont {Gale}, \citenamefont {Garc{\'i}a},
  \citenamefont {Junquera}, \citenamefont {Ordej{\'o}n},\ and\ \citenamefont
  {S{\'a}nchez-Portal}}]{siesta}%
  \BibitemOpen
  \bibfield  {author} {\bibinfo {author} {\bibfnamefont {J.~M.}\ \bibnamefont
  {Soler}}, \bibinfo {author} {\bibfnamefont {E.}~\bibnamefont {Artacho}},
  \bibinfo {author} {\bibfnamefont {J.~D.}\ \bibnamefont {Gale}}, \bibinfo
  {author} {\bibfnamefont {A.}~\bibnamefont {Garc{\'i}a}}, \bibinfo {author}
  {\bibfnamefont {J.}~\bibnamefont {Junquera}}, \bibinfo {author}
  {\bibfnamefont {P.}~\bibnamefont {Ordej{\'o}n}}, \ and\ \bibinfo {author}
  {\bibfnamefont {D.}~\bibnamefont {S{\'a}nchez-Portal}},\ }\href {\doibase
  10.1088/0953-8984/14/11/302} {\bibfield  {journal} {\bibinfo  {journal} {J.
  Phys. Condens. Matter}\ }\textbf {\bibinfo {volume} {14}},\ \bibinfo {pages}
  {2745} (\bibinfo {year} {2002})}\BibitemShut {NoStop}%
\bibitem [{\citenamefont {Artacho}\ \emph {et~al.}(2008)\citenamefont
  {Artacho}, \citenamefont {Anglada}, \citenamefont {Diéguez}, \citenamefont
  {Gale}, \citenamefont {García}, \citenamefont {Junquera}, \citenamefont
  {Martin}, \citenamefont {Ordejón}, \citenamefont {Pruneda}, \citenamefont
  {Sánchez-Portal},\ and\ \citenamefont {Soler}}]{0953-8984-20-6-064208}%
  \BibitemOpen
  \bibfield  {author} {\bibinfo {author} {\bibfnamefont {E.}~\bibnamefont
  {Artacho}}, \bibinfo {author} {\bibfnamefont {E.}~\bibnamefont {Anglada}},
  \bibinfo {author} {\bibfnamefont {O.}~\bibnamefont {Diéguez}}, \bibinfo
  {author} {\bibfnamefont {J.~D.}\ \bibnamefont {Gale}}, \bibinfo {author}
  {\bibfnamefont {A.}~\bibnamefont {García}}, \bibinfo {author} {\bibfnamefont
  {J.}~\bibnamefont {Junquera}}, \bibinfo {author} {\bibfnamefont {R.~M.}\
  \bibnamefont {Martin}}, \bibinfo {author} {\bibfnamefont {P.}~\bibnamefont
  {Ordejón}}, \bibinfo {author} {\bibfnamefont {J.~M.}\ \bibnamefont
  {Pruneda}}, \bibinfo {author} {\bibfnamefont {D.}~\bibnamefont
  {Sánchez-Portal}}, \ and\ \bibinfo {author} {\bibfnamefont {J.~M.}\
  \bibnamefont {Soler}},\ }\href {\doibase 10.1088/0953-8984/20/6/064208}
  {\bibfield  {journal} {\bibinfo  {journal} {J. Phys. Condens. Matter}\
  }\textbf {\bibinfo {volume} {20}},\ \bibinfo {pages} {064208} (\bibinfo
  {year} {2008})}\BibitemShut {NoStop}%
\bibitem [{\citenamefont {Froyen}\ \emph {et~al.}()\citenamefont {Froyen},
  \citenamefont {Troullier}, \citenamefont {Martins}, \citenamefont {Balb\'as},
  \citenamefont {Soler},\ and\ \citenamefont {Garcia}}]{atom}%
  \BibitemOpen
  \bibfield  {author} {\bibinfo {author} {\bibfnamefont {S.}~\bibnamefont
  {Froyen}}, \bibinfo {author} {\bibfnamefont {N.~J.}\ \bibnamefont
  {Troullier}}, \bibinfo {author} {\bibfnamefont {J.~L.}\ \bibnamefont
  {Martins}}, \bibinfo {author} {\bibfnamefont {L.~C.}\ \bibnamefont
  {Balb\'as}}, \bibinfo {author} {\bibfnamefont {J.~M.}\ \bibnamefont {Soler}},
  \ and\ \bibinfo {author} {\bibfnamefont {A.}~\bibnamefont {Garcia}},\
  }\href@noop {} {\enquote {\bibinfo {title} {{ATOM}, a program for {DFT}
  calculations in atoms and pseudopotential generation},}\ }\bibinfo
  {howpublished}
  {\url{https://departments.icmab.es/leem/siesta/Pseudopotentials/Code/downloads.html}}\BibitemShut
  {NoStop}%
\bibitem [{\citenamefont {Troullier}\ and\ \citenamefont
  {Martins}(1991)}]{troullier}%
  \BibitemOpen
  \bibfield  {author} {\bibinfo {author} {\bibfnamefont {N.}~\bibnamefont
  {Troullier}}\ and\ \bibinfo {author} {\bibfnamefont {J.~L.}\ \bibnamefont
  {Martins}},\ }\href {\doibase 10.1103/PhysRevB.43.1993} {\bibfield  {journal}
  {\bibinfo  {journal} {Phys. Rev. B}\ }\textbf {\bibinfo {volume} {43}},\
  \bibinfo {pages} {1993} (\bibinfo {year} {1991})}\BibitemShut {NoStop}%
\bibitem [{\citenamefont {Ullah}\ \emph {et~al.}(2018)\citenamefont {Ullah},
  \citenamefont {Artacho},\ and\ \citenamefont
  {Correa}}]{PhysRevLett.121.116401}%
  \BibitemOpen
  \bibfield  {author} {\bibinfo {author} {\bibfnamefont {R.}~\bibnamefont
  {Ullah}}, \bibinfo {author} {\bibfnamefont {E.}~\bibnamefont {Artacho}}, \
  and\ \bibinfo {author} {\bibfnamefont {A.~A.}\ \bibnamefont {Correa}},\
  }\href {\doibase 10.1103/PhysRevLett.121.116401} {\bibfield  {journal}
  {\bibinfo  {journal} {Phys. Rev. Lett.}\ }\textbf {\bibinfo {volume} {121}},\
  \bibinfo {pages} {116401} (\bibinfo {year} {2018})}\BibitemShut {NoStop}%
\bibitem [{\citenamefont {Ceperley}\ and\ \citenamefont
  {Alder}(1980)}]{ceperley}%
  \BibitemOpen
  \bibfield  {author} {\bibinfo {author} {\bibfnamefont {D.~M.}\ \bibnamefont
  {Ceperley}}\ and\ \bibinfo {author} {\bibfnamefont {B.~J.}\ \bibnamefont
  {Alder}},\ }\href {\doibase 10.1103/PhysRevLett.45.566} {\bibfield  {journal}
  {\bibinfo  {journal} {Phys. Rev. Lett.}\ }\textbf {\bibinfo {volume} {45}},\
  \bibinfo {pages} {566} (\bibinfo {year} {1980})}\BibitemShut {NoStop}%
\bibitem [{\citenamefont {Perdew}\ and\ \citenamefont
  {Zunger}(1981)}]{PhysRevB.23.5048}%
  \BibitemOpen
  \bibfield  {author} {\bibinfo {author} {\bibfnamefont {J.~P.}\ \bibnamefont
  {Perdew}}\ and\ \bibinfo {author} {\bibfnamefont {A.}~\bibnamefont
  {Zunger}},\ }\href {\doibase 10.1103/PhysRevB.23.5048} {\bibfield  {journal}
  {\bibinfo  {journal} {Phys. Rev. B}\ }\textbf {\bibinfo {volume} {23}},\
  \bibinfo {pages} {5048} (\bibinfo {year} {1981})}\BibitemShut {NoStop}%
\bibitem [{\citenamefont {Kohn}\ and\ \citenamefont {Sham}(1965)}]{kohnsham}%
  \BibitemOpen
  \bibfield  {author} {\bibinfo {author} {\bibfnamefont {W.}~\bibnamefont
  {Kohn}}\ and\ \bibinfo {author} {\bibfnamefont {L.~J.}\ \bibnamefont
  {Sham}},\ }\href {\doibase 10.1103/PhysRev.140.A1133} {\bibfield  {journal}
  {\bibinfo  {journal} {Phys. Rev.}\ }\textbf {\bibinfo {volume} {140}},\
  \bibinfo {pages} {A1133} (\bibinfo {year} {1965})}\BibitemShut {NoStop}%
\bibitem [{\citenamefont {Runge}\ and\ \citenamefont {Gross}(1984)}]{runge}%
  \BibitemOpen
  \bibfield  {author} {\bibinfo {author} {\bibfnamefont {E.}~\bibnamefont
  {Runge}}\ and\ \bibinfo {author} {\bibfnamefont {E.~K.~U.}\ \bibnamefont
  {Gross}},\ }\href {\doibase 10.1103/PhysRevLett.52.997} {\bibfield  {journal}
  {\bibinfo  {journal} {Phys. Rev. Lett.}\ }\textbf {\bibinfo {volume} {52}},\
  \bibinfo {pages} {997} (\bibinfo {year} {1984})}\BibitemShut {NoStop}%
\bibitem [{\citenamefont {Pietsch}\ \emph {et~al.}(1976)\citenamefont
  {Pietsch}, \citenamefont {Hauser},\ and\ \citenamefont
  {Neuwirth}}]{PIETSCH197679}%
  \BibitemOpen
  \bibfield  {author} {\bibinfo {author} {\bibfnamefont {W.}~\bibnamefont
  {Pietsch}}, \bibinfo {author} {\bibfnamefont {U.}~\bibnamefont {Hauser}}, \
  and\ \bibinfo {author} {\bibfnamefont {W.}~\bibnamefont {Neuwirth}},\ }\href
  {\doibase https://doi.org/10.1016/0029-554X(76)90714-X} {\bibfield  {journal}
  {\bibinfo  {journal} {Nucl. Instrum. Methods}\ }\textbf {\bibinfo {volume}
  {132}},\ \bibinfo {pages} {79 } (\bibinfo {year} {1976})}\BibitemShut
  {NoStop}%
\bibitem [{\citenamefont {Tomfohr}\ and\ \citenamefont
  {Sankey}()}]{doi:10.1002/1521-3951}%
  \BibitemOpen
  \bibfield  {author} {\bibinfo {author} {\bibfnamefont {J.~K.}\ \bibnamefont
  {Tomfohr}}\ and\ \bibinfo {author} {\bibfnamefont {O.~F.}\ \bibnamefont
  {Sankey}},\ }\href {\doibase
  10.1002/1521-3951(200107)226:1<115::AID-PSSB115>3.0.CO;2-5} {\bibfield
  {journal} {\bibinfo  {journal} {Phys. Status Solidi B}\ }\textbf {\bibinfo
  {volume} {226}},\ \bibinfo {pages} {115}}\BibitemShut {NoStop}%
\bibitem [{\citenamefont {Artacho}\ and\ \citenamefont
  {O'Regan}(2017)}]{PhysRevB.95.115155}%
  \BibitemOpen
  \bibfield  {author} {\bibinfo {author} {\bibfnamefont {E.}~\bibnamefont
  {Artacho}}\ and\ \bibinfo {author} {\bibfnamefont {D.~D.}\ \bibnamefont
  {O'Regan}},\ }\href {\doibase 10.1103/PhysRevB.95.115155} {\bibfield
  {journal} {\bibinfo  {journal} {Phys. Rev. B}\ }\textbf {\bibinfo {volume}
  {95}},\ \bibinfo {pages} {115155} (\bibinfo {year} {2017})}\BibitemShut
  {NoStop}%
\bibitem [{\citenamefont {Deasy}(1994)}]{doi:10.1118/1.597176}%
  \BibitemOpen
  \bibfield  {author} {\bibinfo {author} {\bibfnamefont {J.}~\bibnamefont
  {Deasy}},\ }\href {\doibase 10.1118/1.597176} {\bibfield  {journal} {\bibinfo
   {journal} {Med. Phys.}\ }\textbf {\bibinfo {volume} {21}},\ \bibinfo {pages}
  {709} (\bibinfo {year} {1994})}\BibitemShut {NoStop}%
\bibitem [{\citenamefont {Wang}\ \emph {et~al.}(1998)\citenamefont {Wang},
  \citenamefont {Nagy},\ and\ \citenamefont {Echenique}}]{PhysRevB.58.2357}%
  \BibitemOpen
  \bibfield  {author} {\bibinfo {author} {\bibfnamefont {N.-P.}\ \bibnamefont
  {Wang}}, \bibinfo {author} {\bibfnamefont {I.}~\bibnamefont {Nagy}}, \ and\
  \bibinfo {author} {\bibfnamefont {P.~M.}\ \bibnamefont {Echenique}},\ }\href
  {\doibase 10.1103/PhysRevB.58.2357} {\bibfield  {journal} {\bibinfo
  {journal} {Phys. Rev. B}\ }\textbf {\bibinfo {volume} {58}},\ \bibinfo
  {pages} {2357} (\bibinfo {year} {1998})}\BibitemShut {NoStop}%
\bibitem [{\citenamefont {Wang}\ and\ \citenamefont
  {Nagy}(1997)}]{PhysRevA.56.4795}%
  \BibitemOpen
  \bibfield  {author} {\bibinfo {author} {\bibfnamefont {N.-P.}\ \bibnamefont
  {Wang}}\ and\ \bibinfo {author} {\bibfnamefont {I.}~\bibnamefont {Nagy}},\
  }\href@noop {} {\bibfield  {journal} {\bibinfo  {journal} {Phys. Rev. A}\
  }\textbf {\bibinfo {volume} {56}},\ \bibinfo {pages} {4795} (\bibinfo {year}
  {1997})}\BibitemShut {NoStop}%
\bibitem [{\citenamefont {Winter}\ \emph {et~al.}(2003)\citenamefont {Winter},
  \citenamefont {Juaristi}, \citenamefont {Nagy}, \citenamefont {Arnau},\ and\
  \citenamefont {Echenique}}]{PhysRevB.67.245401}%
  \BibitemOpen
  \bibfield  {author} {\bibinfo {author} {\bibfnamefont {H.}~\bibnamefont
  {Winter}}, \bibinfo {author} {\bibfnamefont {J.~I.}\ \bibnamefont
  {Juaristi}}, \bibinfo {author} {\bibfnamefont {I.}~\bibnamefont {Nagy}},
  \bibinfo {author} {\bibfnamefont {A.}~\bibnamefont {Arnau}}, \ and\ \bibinfo
  {author} {\bibfnamefont {P.~M.}\ \bibnamefont {Echenique}},\ }\href {\doibase
  10.1103/PhysRevB.67.245401} {\bibfield  {journal} {\bibinfo  {journal} {Phys.
  Rev. B}\ }\textbf {\bibinfo {volume} {67}},\ \bibinfo {pages} {245401}
  (\bibinfo {year} {2003})}\BibitemShut {NoStop}%
\bibitem [{\citenamefont {Lindhard}(1965)}]{lindhard1965}%
  \BibitemOpen
  \bibfield  {author} {\bibinfo {author} {\bibfnamefont {J.}~\bibnamefont
  {Lindhard}},\ }\href@noop {} {\bibfield  {journal} {\bibinfo  {journal}
  {Matematisk-fysiske Meddelelser}\ }\textbf {\bibinfo {volume} {14}} (\bibinfo
  {year} {1965})}\BibitemShut {NoStop}%
\bibitem [{\citenamefont {Elman}\ \emph {et~al.}(1984)\citenamefont {Elman},
  \citenamefont {Braunstein}, \citenamefont {Dresselhaus}, \citenamefont
  {Dresselhaus}, \citenamefont {Venkatesan},\ and\ \citenamefont
  {Wilkens}}]{doi:10.1063/1.334210}%
  \BibitemOpen
  \bibfield  {author} {\bibinfo {author} {\bibfnamefont {B.~S.}\ \bibnamefont
  {Elman}}, \bibinfo {author} {\bibfnamefont {G.}~\bibnamefont {Braunstein}},
  \bibinfo {author} {\bibfnamefont {M.~S.}\ \bibnamefont {Dresselhaus}},
  \bibinfo {author} {\bibfnamefont {G.}~\bibnamefont {Dresselhaus}}, \bibinfo
  {author} {\bibfnamefont {T.}~\bibnamefont {Venkatesan}}, \ and\ \bibinfo
  {author} {\bibfnamefont {B.}~\bibnamefont {Wilkens}},\ }\href {\doibase
  10.1063/1.334210} {\bibfield  {journal} {\bibinfo  {journal} {J. Appl.
  Phys.}\ }\textbf {\bibinfo {volume} {56}},\ \bibinfo {pages} {2114} (\bibinfo
  {year} {1984})}\BibitemShut {NoStop}%
\bibitem [{\citenamefont {Iwata}\ \emph {et~al.}(1975)\citenamefont {Iwata},
  \citenamefont {Komaki}, \citenamefont {Tomimitsu}, \citenamefont {Kawatsura},
  \citenamefont {Ozawa},\ and\ \citenamefont
  {Doi}}]{doi:10.1080/00337577508239479}%
  \BibitemOpen
  \bibfield  {author} {\bibinfo {author} {\bibfnamefont {T.}~\bibnamefont
  {Iwata}}, \bibinfo {author} {\bibfnamefont {K.-I.}\ \bibnamefont {Komaki}},
  \bibinfo {author} {\bibfnamefont {H.}~\bibnamefont {Tomimitsu}}, \bibinfo
  {author} {\bibfnamefont {K.}~\bibnamefont {Kawatsura}}, \bibinfo {author}
  {\bibfnamefont {K.}~\bibnamefont {Ozawa}}, \ and\ \bibinfo {author}
  {\bibfnamefont {K.}~\bibnamefont {Doi}},\ }\href {\doibase
  10.1080/00337577508239479} {\bibfield  {journal} {\bibinfo  {journal}
  {Radiation Effects}\ }\textbf {\bibinfo {volume} {24}},\ \bibinfo {pages}
  {63} (\bibinfo {year} {1975})}\BibitemShut {NoStop}%
\bibitem [{\citenamefont {Schroyen}\ \emph {et~al.}(1986)\citenamefont
  {Schroyen}, \citenamefont {Bruggeman}, \citenamefont {Dezsi},\ and\
  \citenamefont {Langouche}}]{SCHROYEN1986341}%
  \BibitemOpen
  \bibfield  {author} {\bibinfo {author} {\bibfnamefont {D.}~\bibnamefont
  {Schroyen}}, \bibinfo {author} {\bibfnamefont {M.}~\bibnamefont {Bruggeman}},
  \bibinfo {author} {\bibfnamefont {I.}~\bibnamefont {Dezsi}}, \ and\ \bibinfo
  {author} {\bibfnamefont {G.}~\bibnamefont {Langouche}},\ }\href {\doibase
  https://doi.org/10.1016/0168-583X(86)90316-2} {\bibfield  {journal} {\bibinfo
   {journal} {Nucl. Instrum. Meth. B}\ }\textbf {\bibinfo {volume} {15}},\
  \bibinfo {pages} {341 } (\bibinfo {year} {1986})}\BibitemShut {NoStop}%
\bibitem [{\citenamefont {Figueroa}\ \emph {et~al.}(2007)\citenamefont
  {Figueroa}, \citenamefont {Cantero}, \citenamefont {Eckardt}, \citenamefont
  {Lantschner}, \citenamefont {Vald\'es},\ and\ \citenamefont
  {Arista}}]{PhysRevA.75.010901}%
  \BibitemOpen
  \bibfield  {author} {\bibinfo {author} {\bibfnamefont {E.~A.}\ \bibnamefont
  {Figueroa}}, \bibinfo {author} {\bibfnamefont {E.~D.}\ \bibnamefont
  {Cantero}}, \bibinfo {author} {\bibfnamefont {J.~C.}\ \bibnamefont
  {Eckardt}}, \bibinfo {author} {\bibfnamefont {G.~H.}\ \bibnamefont
  {Lantschner}}, \bibinfo {author} {\bibfnamefont {J.~E.}\ \bibnamefont
  {Vald\'es}}, \ and\ \bibinfo {author} {\bibfnamefont {N.~R.}\ \bibnamefont
  {Arista}},\ }\href {\doibase 10.1103/PhysRevA.75.010901} {\bibfield
  {journal} {\bibinfo  {journal} {Phys. Rev. A}\ }\textbf {\bibinfo {volume}
  {75}},\ \bibinfo {pages} {010901(R)} (\bibinfo {year} {2007})}\BibitemShut
  {NoStop}%
\bibitem [{\citenamefont {Markin}\ \emph {et~al.}(2008)\citenamefont {Markin},
  \citenamefont {Primetzhofer}, \citenamefont {Prusa}, \citenamefont
  {Brunmayr}, \citenamefont {Kowarik}, \citenamefont {Aumayr},\ and\
  \citenamefont {Bauer}}]{PhysRevB.78.195122}%
  \BibitemOpen
  \bibfield  {author} {\bibinfo {author} {\bibfnamefont {S.~N.}\ \bibnamefont
  {Markin}}, \bibinfo {author} {\bibfnamefont {D.}~\bibnamefont
  {Primetzhofer}}, \bibinfo {author} {\bibfnamefont {S.}~\bibnamefont {Prusa}},
  \bibinfo {author} {\bibfnamefont {M.}~\bibnamefont {Brunmayr}}, \bibinfo
  {author} {\bibfnamefont {G.}~\bibnamefont {Kowarik}}, \bibinfo {author}
  {\bibfnamefont {F.}~\bibnamefont {Aumayr}}, \ and\ \bibinfo {author}
  {\bibfnamefont {P.}~\bibnamefont {Bauer}},\ }\href {\doibase
  10.1103/PhysRevB.78.195122} {\bibfield  {journal} {\bibinfo  {journal} {Phys.
  Rev. B}\ }\textbf {\bibinfo {volume} {78}},\ \bibinfo {pages} {195122}
  (\bibinfo {year} {2008})}\BibitemShut {NoStop}%
\bibitem [{\citenamefont {Primetzhofer}\ \emph {et~al.}(2011)\citenamefont
  {Primetzhofer}, \citenamefont {Rund}, \citenamefont {Roth}, \citenamefont
  {Goebl},\ and\ \citenamefont {Bauer}}]{PhysRevLett.107.163201}%
  \BibitemOpen
  \bibfield  {author} {\bibinfo {author} {\bibfnamefont {D.}~\bibnamefont
  {Primetzhofer}}, \bibinfo {author} {\bibfnamefont {S.}~\bibnamefont {Rund}},
  \bibinfo {author} {\bibfnamefont {D.}~\bibnamefont {Roth}}, \bibinfo {author}
  {\bibfnamefont {D.}~\bibnamefont {Goebl}}, \ and\ \bibinfo {author}
  {\bibfnamefont {P.}~\bibnamefont {Bauer}},\ }\href {\doibase
  10.1103/PhysRevLett.107.163201} {\bibfield  {journal} {\bibinfo  {journal}
  {Phys. Rev. Lett.}\ }\textbf {\bibinfo {volume} {107}},\ \bibinfo {pages}
  {163201} (\bibinfo {year} {2011})}\BibitemShut {NoStop}%
\bibitem [{\citenamefont {Markin}\ \emph {et~al.}(2009)\citenamefont {Markin},
  \citenamefont {Primetzhofer}, \citenamefont {Spitz},\ and\ \citenamefont
  {Bauer}}]{PhysRevB.80.205105}%
  \BibitemOpen
  \bibfield  {author} {\bibinfo {author} {\bibfnamefont {S.~N.}\ \bibnamefont
  {Markin}}, \bibinfo {author} {\bibfnamefont {D.}~\bibnamefont
  {Primetzhofer}}, \bibinfo {author} {\bibfnamefont {M.}~\bibnamefont {Spitz}},
  \ and\ \bibinfo {author} {\bibfnamefont {P.}~\bibnamefont {Bauer}},\ }\href
  {\doibase 10.1103/PhysRevB.80.205105} {\bibfield  {journal} {\bibinfo
  {journal} {Phys. Rev. B}\ }\textbf {\bibinfo {volume} {80}},\ \bibinfo
  {pages} {205105} (\bibinfo {year} {2009})}\BibitemShut {NoStop}%
\bibitem [{\citenamefont {http://www.hpc.cam.ac.uk}()}]{csd3}%
  \BibitemOpen
  \bibfield  {author} {\bibinfo {author} {\bibnamefont
  {http://www.hpc.cam.ac.uk}},\ }\href@noop {} {}\BibitemShut {NoStop}%
\bibitem [{\citenamefont {www.dirac.ac.uk}()}]{dirac}%
  \BibitemOpen
  \bibfield  {author} {\bibinfo {author} {\bibnamefont {www.dirac.ac.uk}},\
  }\href@noop {} {}\BibitemShut {NoStop}%
\bibitem [{\citenamefont {Junquera}\ \emph {et~al.}(2001)\citenamefont
  {Junquera}, \citenamefont {Paz}, \citenamefont {S\'anchez-Portal},\ and\
  \citenamefont {Artacho}}]{junquera}%
  \BibitemOpen
  \bibfield  {author} {\bibinfo {author} {\bibfnamefont {J.}~\bibnamefont
  {Junquera}}, \bibinfo {author} {\bibfnamefont {O.}~\bibnamefont {Paz}},
  \bibinfo {author} {\bibfnamefont {D.}~\bibnamefont {S\'anchez-Portal}}, \
  and\ \bibinfo {author} {\bibfnamefont {E.}~\bibnamefont {Artacho}},\ }\href
  {\doibase 10.1103/PhysRevB.64.235111} {\bibfield  {journal} {\bibinfo
  {journal} {Phys. Rev. B}\ }\textbf {\bibinfo {volume} {64}},\ \bibinfo
  {pages} {235111} (\bibinfo {year} {2001})}\BibitemShut {NoStop}%
\end{thebibliography}%

\end{document}